\documentclass[twocolumn,amsmath,amssymb,superscriptaddress,longbibliography,aps, prb,floatfix]{revtex4-2}

\usepackage{graphicx}% Include figure files
\usepackage{dcolumn}% Align table columns on decimal point
\usepackage{bm}% bold math
\usepackage[dvipsnames]{xcolor}
\usepackage{hyperref}

\usepackage{array}     % For better table formatting
\usepackage{booktabs}  % For professional-looking tables
\usepackage{float}     % For [H] float placement

\usepackage{amsmath}
\allowdisplaybreaks % <----------

\newcommand{\vk}{\mathbf{k}}
\newcommand{\vq}{\mathbf{q}}

\newcommand{\ts}{TaS$_2$}
\newcommand{\mathdefault}[1]{#1}
\newcommand{\nb}{NbSe$_2$}

\begin{document}

%\preprint{APS/123-QED}

\title{Unconventional superconductivity in monolayer transition metal dichalcogenides}
%Force line breaks with \\

\author{Subhojit Roy}
\affiliation{Department of Physics, Indian Institute of Technology Madras, Chennai, 600036, India}
\affiliation{Center for Atomistic Modelling and Materials Design, IIT Madras, Chennai 600036, India}
\affiliation{Quantum Centers in Diamond and Emergent Materials (QCenDiem)-Group, IIT Madras, Chennai, 600036 India}
\author{Andreas Kreisel}
\affiliation{Department of Physics, Indian Institute of Technology Madras, Chennai, 600036, India}
\affiliation{Quantum Centers in Diamond and Emergent Materials (QCenDiem)-Group, IIT Madras, Chennai, 600036 India}
\affiliation{Niels Bohr Institute, University of Copenhagen, DK-2100 Copenhagen, Denmark}
\author{Brian M. Andersen}
\affiliation{Niels Bohr Institute, University of Copenhagen, DK-2100 Copenhagen, Denmark}
\author{Shantanu Mukherjee}
\email[Corresponding Author: ]{shantanu@iitm.ac.in}
\affiliation{Department of Physics, Indian Institute of Technology Madras, Chennai, 600036, India}
\affiliation{Center for Atomistic Modelling and Materials Design, IIT Madras, Chennai 600036, India}
\affiliation{Quantum Centers in Diamond and Emergent Materials (QCenDiem)-Group, IIT Madras, Chennai, 600036 India}

\begin{abstract}
A variety of experimental observations in monolayer transition metal dichalcogenide superconductors with Ising spin-orbit coupling suggest the presence of an unconventional superconducting pairing mechanism. Some of these experiments include observation of Leggett modes and a nodal superconducting gap in STM experiments, a large in-plane upper critical field compared to the Pauli limit, and the observation of a twofold gap anisotropy in magnetoresistance measurements.  Here, we propose a superconducting pairing mechanism mediated by spin and charge fluctuations and identify the dominant superconducting instability relevant to monolayer TaS$_2$. We then explore the effect of an additional electron-phonon pairing contribution, and compare our results with recent experimental findings. In particular, our theory stabilizes a superconducting ground state with nodal-like density of states that agrees with STM experiments. The theory obtains a large in-plane upper critical field due to a combination of Ising spin-orbit coupling and even-odd parity mixing in the superconducting state. Further, we find that an in-plane magnetic field splits the degeneracy of the superconducting ground state, and the resulting twofold symmetric superconducting order parameter could explain the gap anisotropy observed in magnetoresistance experiments. Overall, the proposed theoretical pairing model can reconcile diverse experimental observations and remains consistent with observations on other dichalcogenide superconductors such as monolayer NbSe$_2$.

\end{abstract}

\maketitle

\section{Introduction}
Transition metal dichalcogenides (TMDs) as quasi two-dimensional (2D) materials are known to host unique electronic, optical, and mechanical properties that diverge markedly from their bulk counterparts. In bulk TMDs superconductivity is mediated by conventional electron-phonon interaction~\cite{frindt1972superconductivity,kobayashi1977thermodynamic,garoche1978dynamic,staley2009electric,el2013superconductivity,CLAYMAN19711881,Boaknin2003,huang2007experimental,valla2004quasiparticle, Noat2015}, whereas there have been reports of unconventional superconductivity in their monolayer limit~\cite{galvis2014zero,wan2022observation,ganguli2022confinement,wickramaratne2020ising,fatemi2018electrically,yuan2014possible,roldan2013interactions,khezerlou2016transport,hsu2017topological,he2018magnetic}. Such monolayer TMDs are believed to contain enhanced spin and charge fluctuations compared to their bulk counterparts, resulting from reduced screening of electronic correlations as a result of the dimensional reduction. In the presence of stronger electronic correlations, the superconducting state can stabilize an unconventional gap function that supports chiral and nodal superconductivity.~\cite{silber2024two,siegl2025friedel,bh1n-sc95}. 

\begin{figure}
\centering
    \includegraphics[width=1.0\linewidth]{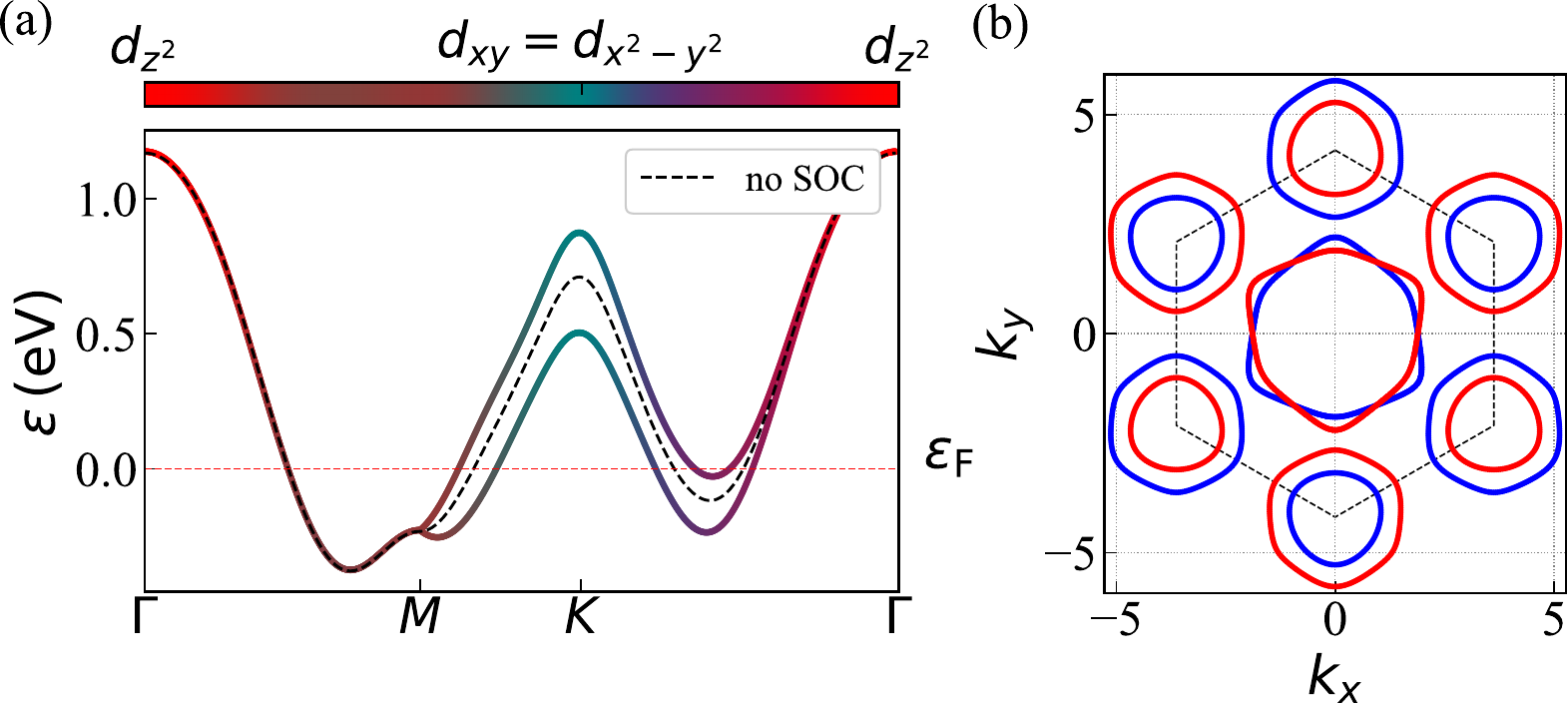}
 \caption{(a) shows the low-energy orbital-resolved electronic structure of monolayer TaS$_2$. The black dashed line denotes the low-energy band in the absence of Ising SOC. (b) presents the spin-resolved Fermi surface in the presence of Ising SOC. The results in both panels are obtained from a tight-binding model derived from relativistic DFT calculations.}
    \label{fig_1}
 \end{figure}

The 1H-TMDs are composed of transition metal atoms (e.g., Nb, Ta, W) sandwiched between chalcogen atoms (Se, S, Te) in an MX$_2$ stoichiometry~\cite{Wilson01051969,manzeli20172d}. When reduced to a monolayer, 1H-TMDs like TaS$_2$ or NbSe$_2$ lose their inversion center while retaining an in-plane mirror symmetry, giving rise to a strong antisymmetric spin-orbit coupling (SOC) and corresponding splitting of bands [see Fig.~\ref{fig_1}(a)]. This Ising SOC locks the electron spins out of the 2D plane, where a spin-momentum locked spin-up state at the $K$ valley is degenerate with a spin-down state at the $K'$ valley [see Fig.~\ref{fig_1}(b)]. When these materials also undergo a superconducting transition, they are dubbed as Ising superconductors~\cite{xing2017ising,wan2022observation,xi2016ising,dvir2018spectroscopy,mockli2018robust,fischer2023superconductivity,wickramaratne2021magnetism,de2018tuning,lu2015evidence,kim2017quasiparticle,saito2016superconductivity,liu2018interface,li2021printable}. Interest in Ising superconductors has been magnified by experimental observations that hinted towards the presence of an unconventional superconducting state. These include an enhanced upper critical field that significantly surpasses the Pauli paramagnetic limit~\cite{lu2015evidence,xi2016ising,lu2018full}, non-monotonic dependence of superconducting transition temperature on disorder concentration~\cite{Zhao2019,Peng2018}, observation of a zero-bias peak in tunneling conductance ~\cite{galvis2014zero}, and the observation of a twofold symmetric gap structure in magnetoresistance measurements~ \cite{hamill2021two,Cho2022,liu2024nematic}. These experiments support a scenario where the superconducting gap is nodal, changes sign, and mixes odd- and even-parity superconducting states in the presence of broken inversion symmetry~\cite{mockli2019magnetic}. 

Ising superconductivity in monolayer 1H-TaS$_2$ ~\cite{de2018tuning,li2024beyond,chen2019topological,lian2022intrinsic} occurs below a critical temperature $T_c \sim 3K$~\cite{ma2018unusual,pan2017enhanced,Peng2018,Bekaert2020}. There have been very few studies addressing its pairing mechanism and possible ground state. A previous work based on symmetry analysis, considering only the d$_{z^{2}}$ orbital, suggested an (s+f)-wave superconducting ground state~\cite{chen2019topological}. The pairing mechanism of another TMD family member, 1H-NbSe$_2$—which features a weaker Ising SOC compared to 1H-TaS$_2$~\cite{de2018tuning}, has been more extensively studied. Various theoretical proposals have considered electron-phonon-mediated pairing~\cite{das2023electron,zheng2019electron,xi2015strongly,wang2023decisive,lian2018unveiling}, while others attribute superconductivity to purely repulsive interactions~\cite{das2021quantitative,costa2022ising,hsu2017topological,shaffer2020crystalline,Ammar2023,Shaffer2023,Horhold_2023,siegl2025friedel}. Some models propose a combination of electron-phonon coupling and spin-fluctuation-mediated pairing~\cite{das2023electron}. The presence of strong electron correlations in monolayer TMDs is indicated by studies of their magnetic properties~\cite{roldan2013interactions,NavarroMoratalla2016} that find further evidence of the presence of significant paramagnetic fluctuations ~\cite{das2023electron}. These fluctuations may provide an important contribution to the superconducting pairing interaction, and naturally support a sign-changing gap structure on the Fermi surface. 

In this work, building on the aforementioned theoretical and experimental studies, we have used a comprehensive framework that incorporates Ising SOC and the multiorbital nature of the electronic structure, under the assumption that superconducting pairing is driven by spin and charge fluctuations~\cite{berk1966effect,scalapino1999superconductivity,Romer2015,Kreisel2017,Romer2019,Kreisel2022}. Using this framework, we derive the ground state superconducting gap structure in TaS$_2$. Superconducting pairing interaction will be influenced by spin-orbit coupling  strength and the location of the paramagnetic susceptibility peak. We find that TaS$_2$ has a larger Ising spin orbit coupling ($\Delta_{SO}\sim 250$meV) compared to monolayer NbSe$_2$ ($\Delta_{SO}\sim130$meV), and the susceptibility peak is also shifted to the momentum transfer $\vq_{\mathrm{max}}=0.34\Gamma\text{--}M$ (Fig. 2)  compared to $\vq_{\mathrm{max}}=0.38\Gamma\text{--}M$ in NbSe$_2$\cite{roy2024unconventional}. Note that we here give the maximum spin splitting to quantify the spin-orbit coupling while Ref. \cite{de2018tuning} quantifies with the average spin splitting yielding a trivial factor of two. However, the dominant superconducting instability remains as the two dimensional E$^{'}$ irreducible representation (Irrep) in monolayer TaS$_2$, similar to our previous study on 1H-NbSe$_2$~\cite{roy2024unconventional}. We additionally find that the obtained superconducting gap structure in TaS$_2$ is consistent with a number of recent experiments including observation of nodal local density of states~\cite{Vano2023,Cho2022} in STM experiments, two fold symmetric gap structure observed in magnetoresistance measurements~\cite{hamill2021two,Cho2022}, and large in-plane upper critical field compared to the Pauli limit that is larger in monolayer TaS$_2$ compared to values obtained in monolayer NbSe$_2$~\cite{de2018tuning}. We also investigate the interplay between spin-fluctuation pairing and an attractive conventional electron-phonon pairing interaction in governing the superconducting gap structure.
%%%%%%%%%%%%%%%%%%%%%%%%%%%%%%%%%%%%%%%%%%%%%%%%%%%%%%%%%%%%%%%
%%%%%%%%%%%%%%%%%%%%%%%%%%%%%%%%%%%%%%%%%%%%%%%%%%%%%%%%%%%%%%%

This paper is structured as follows: Sec. \ref{sec:level2} introduces the applied model Hamiltonian. In Sec. \ref{sec:level3}, we present our numerical results and discuss the implications of Ising SOC. Finally, Sec. \ref{sec:level4} provides a summary and conclusion. Additional technical details can be found in the Appendix.

\section{Model : Microscopic Hamiltonian} \label{sec:level2}
We employ a multiorbital tight-binding Hamiltonian to accurately describe the low-energy electronic structure of monolayer \ts{}, see Appendix \ref{app_A} for details. The relevant basis for this Hamiltonian consists of three Ta \(d\)-orbitals: \(d_{z^2}\), \(d_{x^2 - y^2}\), and \(d_{xy}\).  The Hamiltonian is given by
\begin{eqnarray}  
  H_{\rm TB}=\sum_{\vk l_1 l_2 \sigma} h_{l_1 l_2 \sigma}(\vk)c^{\dagger}_{\vk l_1 \sigma}c_{\vk l_2 \sigma}. \label{eq_tb}  
\end{eqnarray}  
Here, \(c^{\dagger}_{\vk l \sigma}\) creates an electron with spin \(\sigma\), orbital \(l\), and wave vector \(\vk\). In the presence of Ising SOC, the eigenstates of \(S_z\) remain good spin quantum numbers since no spin-flip terms appear in the Hamiltonian.

The above Hamiltonian can be represented in the band basis by performing a unitary transformation
\begin{align}
 c_{\vk,\tilde{l}}=\sum_{\alpha}u^{\alpha}_{\tilde{l}}(\vk)a_{\vk \alpha \sigma},
\label{eq_unitary}
\end{align}
where $u^{\alpha}_{\tilde{l}}(\vk)$ are components of the eigenvector with eigenenergy $\epsilon_\alpha(\vk)$ corresponding to band $\alpha$ and we have used a compact notation $[\tilde{l}:=(l,\sigma)]$.
The low-energy electronic structure yields a single electronic band forming the Fermi surface that splits into two spin momentum locked bands in the presence of Ising SOC. Note that the SOC is not used as a tuning parameter in our study, but the two electronic structures considered in this work: with SOC and without SOC are both derived from DFT calculations as described above. See Appendix A for more details on the electronic structure. 

As shown in Fig.~\ref{fig_1}(a), the two low-energy bands forming the Fermi surface can be described by the Kramers pair $|\vk,\uparrow\rangle$,
and $|-\vk,\downarrow\rangle$.
Both bands contain contributions from all three $d$-orbitals, and each band is described additionally by a particular spin eigenstate due to spin momentum locking. In Fig.~\ref{fig_band_fs}(b) [see Appendix A] we show the Fermi surface together with its orbital content. The $d_{z^2}$ orbital dominates the orbital content of the $\Gamma$-centered Fermi pockets, whereas the $K$ and $K'$ centered pockets are dominated by the $d_{x^2-y^2}$ and $d_{xy}$ orbitals.

The bare electronic repulsion is given by the usual on-site Hubbard-Hund interaction, parametrized by $U$ and $J$. In a compact notation, this interaction takes the form [see Appendix \ref{app_B} for more details]
\begin{eqnarray}
H_{\rm int}=\frac{1}{2} \sum_{\vk,\vk',\tilde{l}} [\hat{V}]^{\tilde{l}_1,\tilde{l}_2}_{\tilde{l}_3,\tilde{l}_4}(\vk,\vk') \, c^{\dagger}_{\vk,\tilde{l}_1} \,c^{\dagger}_{-\vk,\tilde{l}_3} \, c_{-\vk',\tilde{l}_2}\, c_{\vk',\tilde{l}_4}.
\label{eqn:int_hamil}
\end{eqnarray}
Thus the total Hamiltonian applied in the following analysis is given by
 \begin{eqnarray}
     H=H_{\rm TB}+H_{\rm int}\,.\label{eq_ham}
 \end{eqnarray}
\section{Numerical Results and Discussion} \label{sec:level3}
\subsection{Calculation of susceptibility}
Within the random phase approximation (RPA), the bare Coulomb interaction, combined with spin and charge fluctuations, gives rise to an effective interaction that mediates electron pairing, leading to the formation of Cooper pairs. The essential physics of these fluctuations is encapsulated by spin and charge susceptibilities. With the orbital structure taken into account, the non-interacting bare susceptibility tensor at momentum ${\vq}$ is defined as
{\small
\begin{align}
 \chi_{\tilde{l}_3 \tilde{l}_4 }^{\tilde{l}_1 \tilde{l}_2 }(\vq,\tau)\!=\!\!\frac{1}{N}\sum_{\vk,\vk'} & \langle T_{\tau} \, c^{\dagger}_{\vk,\tilde{l}_1}(\tau)c^{}_{\vk+\vq,\tilde{l}_2}(\tau)
 c^{\dagger}_{\vk',\tilde{l}_3}(0)  c_{\vk'-\vq,\tilde{l}_4}(0) \rangle_{0}.
\end{align}
}
Explicitly, this is given by~\cite{Graser2009}
\begin{align}
\chi_{\tilde{l}_3 \tilde{l}_4  }^{\tilde{l}_1 \tilde{l}_2}(\vq,i \omega_{n})&=\frac{1}{N}\sum_{\vk,\alpha,\beta}
\frac{f(\epsilon_{\beta}(\vk+\vq))-f(\epsilon_{\alpha}(\vk))}{i\omega_{n} + \epsilon_{\alpha}(\vk)-\epsilon_{\beta}(\vk+\vq)}
\nonumber\\ &\times
u^{\alpha}_{\tilde{l}_4}(\vk)u^{\alpha *}_{\tilde{l}_1}(\vk)u^{\beta}_{\tilde{l}_2}(\vk+\vq)%\nonumber\\ &\times&
u^{\beta *}_{\tilde{l}_3}(\vk+\vq)
.\label{eq_susceptibility}%\nonumber\\
\end{align}

Here, the eigenenergies \(\epsilon_{\alpha}(\vk)\) are computed at momentum \(\vk\), where \(\alpha = (1, \dots, 6)\) denotes the band index, and \(\vk\) represents points in the Brillouin zone (BZ). The function \( f(x) \) is the Fermi-Dirac distribution, given by  
$f(x) = 1/\left(1 + e^{\beta x}\right),$
where \(\beta = 1 / (k_B T)\) and \( k_B \) is the Boltzmann constant. As mentioned earlier, this expression describes the multiorbital susceptibility. To extract the physical susceptibility, we sum over orbitals, i.e. evaluate
\begin{align}
 \chi^{\sigma_1\sigma_1'}_{\sigma_2\sigma_2'}=\frac{1}{2}\sum_{{l}_{1},{l}_{2}}\chi_{\tilde{l}_{2} \tilde{l}_{2} }^{\tilde{l}_{1} \tilde{l}_{1} }(\vq,0).
\end{align}
Furthermore, the physical susceptibility can be decomposed into the longitudinal ($\chi^{\parallel}$) and transverse ($\chi^{\perp}$) channels  as
\begin{eqnarray}
 \chi^{\parallel}(\vq,\omega)&=&  \frac{1}{4} (\chi_{\uparrow \uparrow }^{\uparrow \uparrow }-\chi_{\downarrow \downarrow }^{\uparrow \uparrow }
 -\chi_{\uparrow \uparrow }^{\downarrow \downarrow }+\chi_{\downarrow \downarrow }^{\downarrow \downarrow }), \\
 \chi^{\perp}(\vq,\omega)&=& \frac{1}{2} (\chi_{\downarrow \uparrow }^{\uparrow \downarrow }+\chi_{\uparrow \downarrow }^{\downarrow \uparrow }),
\end{eqnarray}
where, for ease of notation, we have suppressed the orbital indices that are being summed over. The RPA susceptibility is given by
\begin{align}
    \hat{\chi}^{\mathrm{RPA}} = \frac{\hat{\chi}}{1 - \hat{\chi} \hat{U}},
\end{align}

where \(\hat{U}\) denotes the interaction matrix. Each element of \(\hat{\chi}\) and \(\hat{U}\) is a \(4 \times 4\) matrix in spin space, with every spin-space entry itself being an \(81 \times 81\) matrix in orbital space. Thus, the full tensor structure can be viewed as a \(4 \times 4\) block matrix, where each block is an \(81 \times 81\) matrix.

\begin{figure}
\centering
    \includegraphics[width=1.0\linewidth]{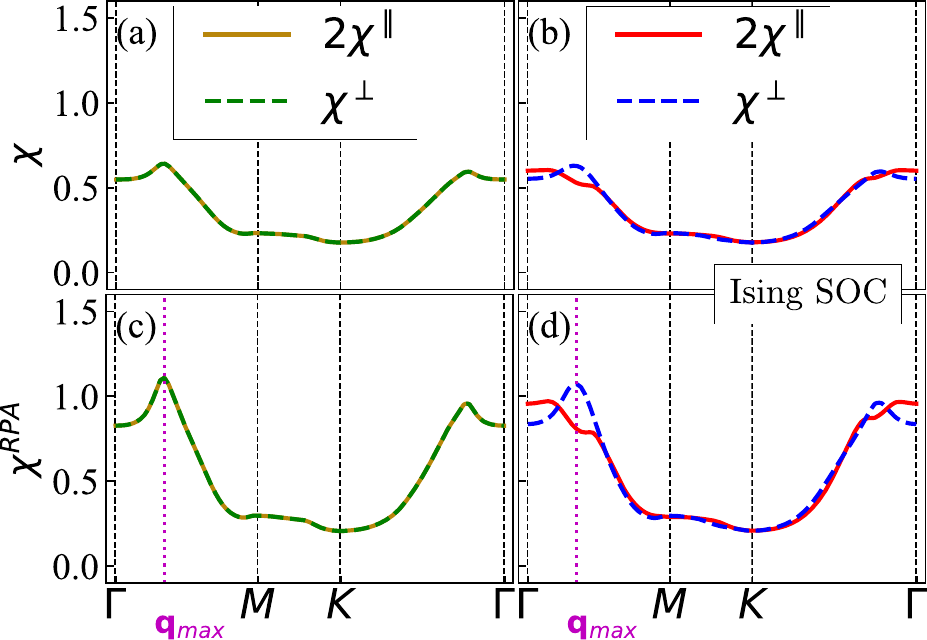}
\caption{Spin susceptibility in units of $1/\mathrm{eV}$ at $\omega=0$ along the high-symmetry path $\mathdefault{\Gamma\text{--}M\text{--}K\text{--}\Gamma}$. (a) and (c) show the noninteracting and corresponding RPA susceptibilities, respectively, calculated without SOC, while (b) and (d) display the corresponding results in the presence of Ising SOC. In both cases, the RPA paramagnetic susceptibility exhibits a peak at a wave vector $\vq_{\mathrm{max}} \sim 0.34\,\Gamma\text{--}M$. For the RPA results we used $U=0.6$\,eV and $J=U/4$.
}
    \label{fig_3}
 \end{figure}
In Fig.~\ref{fig_3}, we present the static physical paramagnetic susceptibility. Figures~\ref{fig_3}(a) and \ref{fig_3}(c) correspond to the noninteracting and RPA susceptibilities in the absence of Ising SOC, respectively, while Figs.~\ref{fig_3}(b) and \ref{fig_3}(d) show the same quantities in the presence of Ising SOC. The inclusion of Ising SOC slightly reduces the noninteracting susceptibility, yet in both cases, the maximum peak appears at \(\vq_{\mathrm{max}} \sim 0.34 \, \Gamma\text{-}M\).  

As expected, Ising SOC breaks the degeneracy between the transverse and longitudinal susceptibility channels. Notably, in the presence of Ising SOC in monolayer TaS$_2$, the transverse susceptibility is enhanced relative to the longitudinal susceptibility, with this effect becoming more pronounced in the interacting RPA susceptibilities. A similar trend is observed in monolayer \nb{}, although the separation between transverse and longitudinal susceptibilities is smaller due to a smaller SOC, and the peak shifts to \(\vq_{\mathrm{max}} \sim 0.38 \, \Gamma\text{-}M\). In the following, we calculate the superconducting pairing kernel from the orbital resolved spin and charge susceptibilities, and solve the linearized BCS gap equation to extract the dominant superconducting instabilities.

\subsection{Leading superconducting instabilities}
Within a spin fluctuation-mediated pairing scenario, the effective superconducting pairing kernel incorporates the mixing of spin and orbital indices induced by the SOC~\cite{Romer2019}. The pairing kernel can be expressed as
\begin{eqnarray}
  [\hat{V}(\vk,\vk')]^{\tilde{l}_1,\tilde{l}_2}_{\tilde{l}_3,\tilde{l}_4}&=&[\hat{U}]^{\tilde{l}_1,\tilde{l}_2}_{\tilde{l}_3,\tilde{l}_4}
  -[\frac{\hat{U}\,\hat{\chi_{}}\,\hat{U}}{1-\hat{\chi_{}}\,\hat{U}}]^{\tilde{l}_1,\tilde{l}_4}_{\tilde{l}_3,\tilde{l}_2}(\vk-\vk') \nonumber \\
  &+&[\frac{\hat{U}\,\hat{\chi_{}}\,\hat{U}}{1-\hat{\chi_{}}\,\hat{U}}]^{\tilde{l}_1,\tilde{l}_2}_{\tilde{l}_3,\tilde{l}_4}(\vk+\vk').
\end{eqnarray}
%%%%%%%%%%%
The interaction matrix $\hat{U}$ and the individual contributions to the pairing kernel are provided in Appendix B.  The superconducting gap function is defined as
\begin{small}
\begin{align}
 \Delta^{\tilde{m}_{1},\tilde{m}_2}(\vk)=-\sum_{\vk',\tilde{m}_{3},\tilde{m}_{4}}V^{\tilde{m}_2,\tilde{m}_1}_{\tilde{m}_3,\tilde{m}_4}(\vk,\vk')\langle a_{\vk,\tilde{m}_3}(\tau)a_{-\vk',\tilde{m}_{4}}(\tau) 
 \rangle,
 \end{align}
\end{small}
where $V_{\tilde{m}_3,\tilde{m}_4}^{\tilde{m}_1,\tilde{m}_2}(\vk,\vk')$ is the pairing interaction projected on the Fermi surface. The indices $[\tilde{m}_i:=(m_i,\sigma_i)]$ represent the band indices and the spin degrees of freedom, respectively; $m_i$ takes the values $(1,2,3)$ and $\sigma_i$ takes the values $(1,2)$. As an example, in this notation the two SOC split bands at the Fermi energy can be represented by the indices $\tilde{m}_1=(1,1)$, and $\tilde{m}_2=(1,2)$.

The linearized gap equation in the presence of Ising SOC can be derived from the Gor'kov Green's functions~\cite{roy2024unconventional}. For a two-band system with bands relevant to monolayer \ts{}, the gap equation can be expressed as
\begin{align}
\Delta^{\tilde{m},\tilde{m}'}_{}(\vk)=-\tilde{I}&\sum_{\substack{ \vk' }}\left[ V^{\tilde{m}',\tilde{m}}_{\tilde{m},\tilde{m}'}(\vk,\vk')
\Delta^{\tilde{m},\tilde{m}'}_{}(\vk') \right.\nonumber\\
 & \left.+   V^{\tilde{m}',\tilde{m}}_{\tilde{m}',\tilde{m}}(\vk,\vk') 
 \Delta^{\tilde{m}',\tilde{m}}_{}(\vk')\right],
 \label{Eq_gap_master}
\end{align}
where $\tilde{I} = \text{ln}\left( \frac{2e^{\gamma}\hbar\omega_{\mathrm{c}}}{\pi k_{B} T_{\mathrm{c}} }\right)$.
The Euler constant is $\gamma\approx 0.5772$ and $\omega_c$ the relevant frequency of the spin fluctuations.
Here the gap equation is restricted to the opposite spin pairing states with $\tilde{m}=(m,\sigma)$, and $\tilde{m}'=(m',\bar{\sigma})$.
A zero energy and a zero momentum gap function $\Delta^{\tilde{m},\tilde{m}'}(\vk)$ pairs opposite spin states, i.e. with different $\tilde m\neq \tilde m'$, but originating from the same $m$. We note that the projection onto the Fermi surface will only give contributions for bands crossing the Fermi level. For the \ts{} monolayer it turns out that only 2 of the 6 bands cross the Fermi level [see Fig. 14(a)] and we therefore only obtain the superconducting order parameters on these two bands.
\subsubsection{Gap solutions without Ising SOC}
\begin{figure}
\centering
    \includegraphics[width=1.0\linewidth]{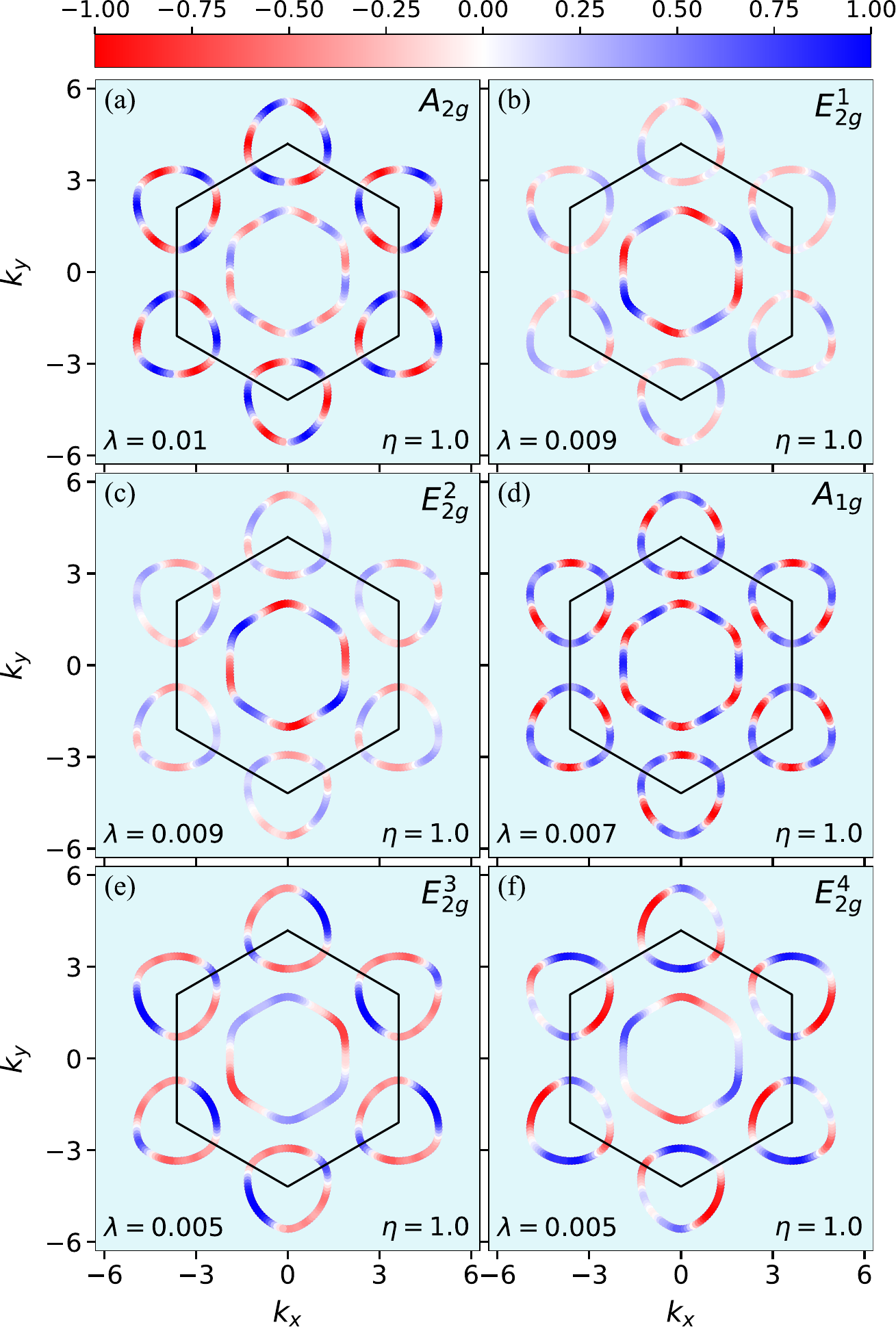}
    \caption{
Solutions of the linearized gap equation $\Delta_{k_{F}}$ plotted on the Fermi surface for the six leading superconducting instabilities with corresponding eigenvalue $\lambda$ and Irreps. Here $\eta = \pm 1$ corresponds to the pure even (odd) parity solutions. The non interacting Hamiltonian used here ignores the Ising SOC. Additional parameters: $U = 0.6$ eV, $J = U/4$.}
    \label{fig_nosoc}
 \end{figure}

Calculations of the superconducting instability arising from spin fluctuations in the absence of SOC have been performed in the literature for various multiband systems~\cite{Graser2009,wang2013superconducting,kemper2010sensitivity,wu2015}. The opposite-spin paired gap function \(\Delta(\mathbf{k})\) of the leading instabilities and the corresponding eigenvalues can be obtained from the linearized gap equation
    \begin{figure}
\centering
    \includegraphics[width=1.0\linewidth]{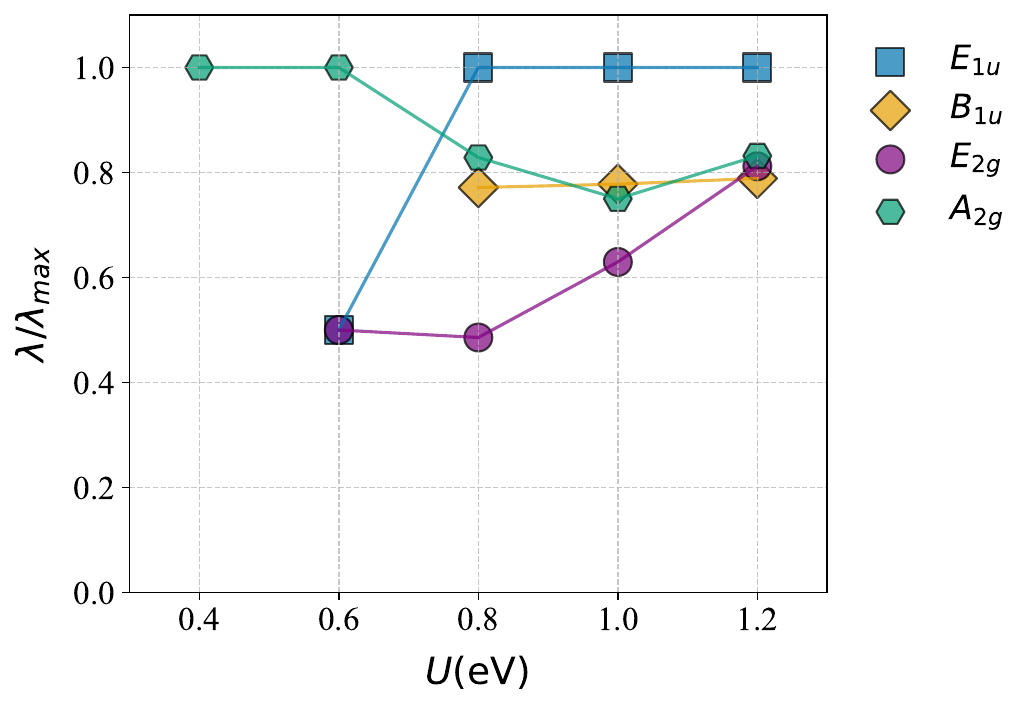}
    \caption{The evolution of the eigenvalues $\lambda/\lambda_{max}$, corresponding to specific Irreps, is presented as a function of $U$ with $J=U/4$ in the absence of Ising SOC. The four largest eigenvalues are plotted. Here, $J = U/4$.}
    \label{fig_irrep_vs_u_no_soc}
 \end{figure}
%%%%%%%%%%%
\begin{align}
\Delta^{m,m'}_{}(\vk)=-\tilde{I}&\sum_{\substack{ \vk' }} V^{m',m}_{m,m'}(\vk,\vk')
\Delta^{m,m'}_{}(\vk').
\end{align}
 %%%%%%%
The solution of the above linearized gap equation is extracted by diagonalizing the $N_k\times N_k$ gap matrix, where $N_k$ is the number of k-points on the Fermi surface. For the problem considered here, we get good convergence with $N_k \sim 600$. We therefore obtain $N_k$ eigenvalues and the corresponding eigenvector components provide us information about the gap function over the Fermi surface. The largest eigenvalue $\lambda$ is a measure of the superconducting transition temperature, and the eigenvector gives the leading pairing instability or superconducting gap function. The other eigenvalues and eigenvectors represent the sub-dominant solutions.

In Fig.~\ref{fig_nosoc}, we present the solutions corresponding to the six largest eigenvalues of the linearized gap equation for \( U = 0.6 \)eV  and $J=U/4$. In the absence of SOC, the underlying symmetry of the gap function is governed by the Irreps of the D$_{6h}$ point group. The leading superconducting instability changes sign across the Fermi surface and belongs to the even parity one-dimensional (1D) A$_{2g}$ irrep [see Fig.~\ref{fig_nosoc}(a)] where the maximum gap magnitude on the Gamma centered pocket is smaller than the corresponding value on the  K-centered pockets. Note that we have additionally used superscripts to represent the Irreps in order to distinguish solutions that belong to the same symmetry class, but have different eigenvalues owing to variations in the nodal structure of the superconducting gap. For example, as shown in Figs.~\ref{fig_nosoc}(b) and \ref{fig_nosoc}(c), $\lambda = 0.009$ gives the first subdominant order that is a degenerate set of solutions E$_{2g}^{1}$, and E$_{2g}^{2}$, whereas $\lambda = 0.005$ contains the fifth subdominant order E$_{2g}^{3}$ and E$_{2g}^{4}$ that belong to the same E$_{2g}$ irrep but differs in their nodal structures.

To gain further insight into how spin fluctuations drive the pairing structure, we conduct a study with \( U \) as a parameter [see Fig.~\ref{fig_irrep_vs_u_no_soc}]. In the weak-coupling RPA approach, the eigenvalues are monotonously increasing with bare (effective) interaction $U$ from very small values to values of unity, so we are showing here and in the following only eigenvalues normalized to the leading eigenvalue to discuss relative competition between superconducting instabilities. Notably, for \( U < 0.8 \) eV, the dominant instabilities favor even-parity A$_{2g}$ symmetry as discussed above, whereas for \( U \geq 0.8 \) eV, the dominant instability switches to the odd-parity E$_{1u}$ symmetry. Similar transition from  A$_{2g}$ symmetry to an  E$_{1u}$ symmetry solution is also obtained in NbSe$_2$ when SOC is ignored ~\cite{roy2024unconventional}, but unlike in TaS$_2$, an E$_{2g}$ symmetry solution also shows up for small  \( U \) values in NbSe$_2$. Additionally, in comparison to NbSe$_2$ we find a relatively larger gap on the Gamma-centered Fermi pocket compared to the gap on the K-centered pocket in TaS$_2$ for the A$_{2g}$ solution.

These differences in gap function solutions between the two TMD family members can be attributed to the nesting properties of the Fermi surface and the corresponding differences in susceptibilities  influencing the structure of the pairing interaction. 
  
\subsubsection{Gap solutions in the presence of Ising SOC}
The lack of inversion symmetry and presence of Ising SOC leads to a complex spin-dependent superconducting pairing interaction that needs to be incorporated in a generalized superconducting gap equation as shown in Eq.~\eqref{Eq_gap_master}. In the normal state, the Ising SOC splits the Fermi surface with each spin-split band being associated with a spin-up or spin down quantum number. Due to this spin momentum locking, in Eq.~\eqref{Eq_gap_master} the band indices $\tilde{m} = (m,\sigma)$ refer to bands with well-defined spin eigenstates $\sigma =(\uparrow,\downarrow)$, respectively. This leads to a larger gap matrix equation, where the gap function pairs electrons from different bands. 
%%%%%%%%%%%%%%%%%%%%%%%%%%
\begin{figure}
\centering
    \includegraphics[width=1.0\linewidth]{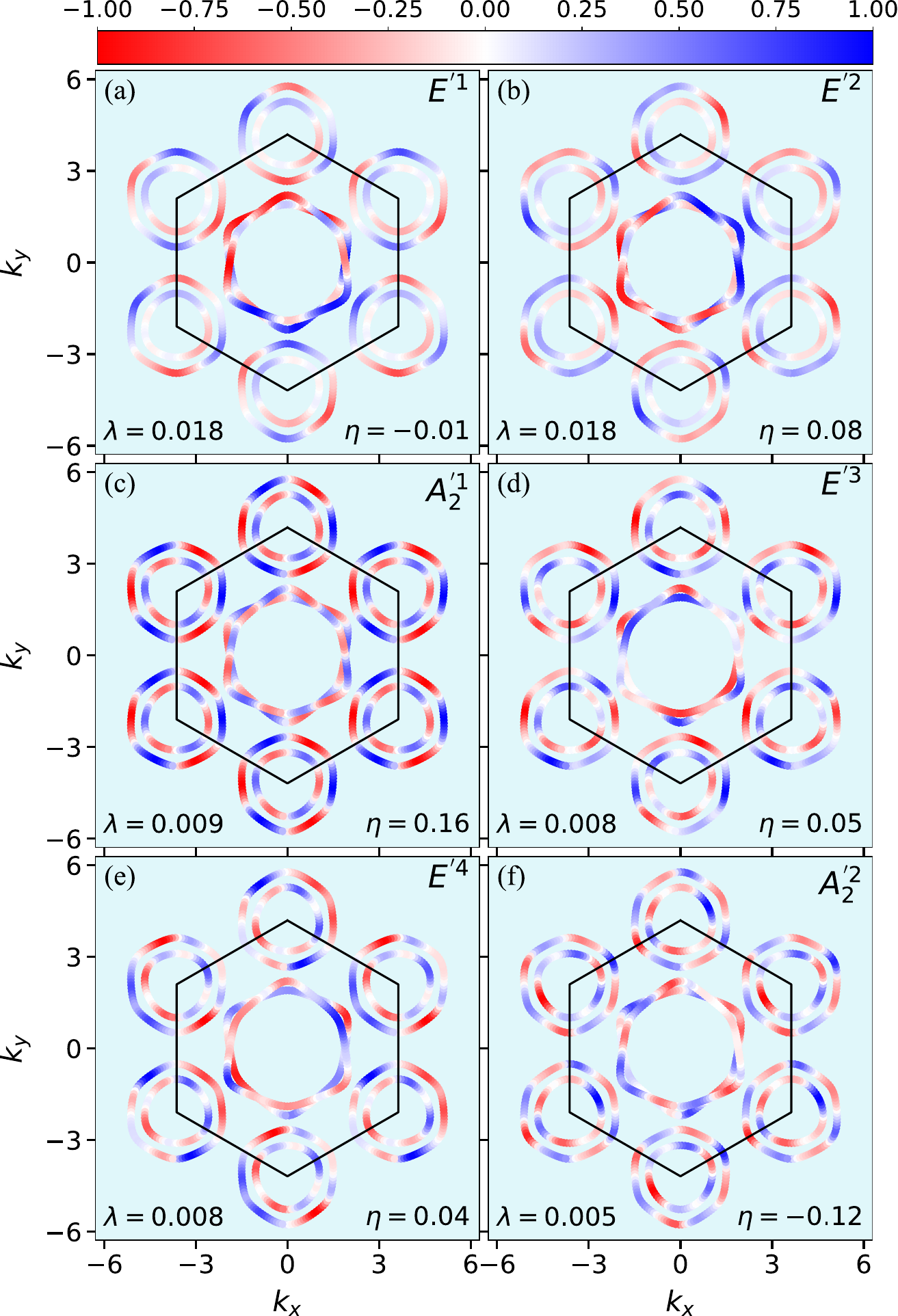}
    \caption{
Superconducting gap functions plotted on Fermi surface in the presence of Ising SOC with $U = 0.6$ eV and $J = U/4$. Here, $\lambda$ on the plots denote the eigenvalues of the corresponding solutions of the gap equation and $\eta$ the even-odd parity mixing. (a) and (b) display the degenerate ground state solution.}
    \label{fig_gap_soc}
 \end{figure}
 %%%%%%%

In the presence of Ising SOC the gap function solutions belong to the irreps of the D$_{3h}$ point group symmetry. In Fig.~\ref{fig_gap_soc} we present the solution for a pairing interaction extracted for $U=0.6$ eV and $J=U/4$. As shown in Fig.~\ref{fig_gap_soc}(a) and \ref{fig_gap_soc}(b), the dominant superconducting instability in the presence of Ising SOC belongs to the two-dimensional E${'}$ irrep. 

\begin{figure}
\centering
    \includegraphics[width=1.0\linewidth]{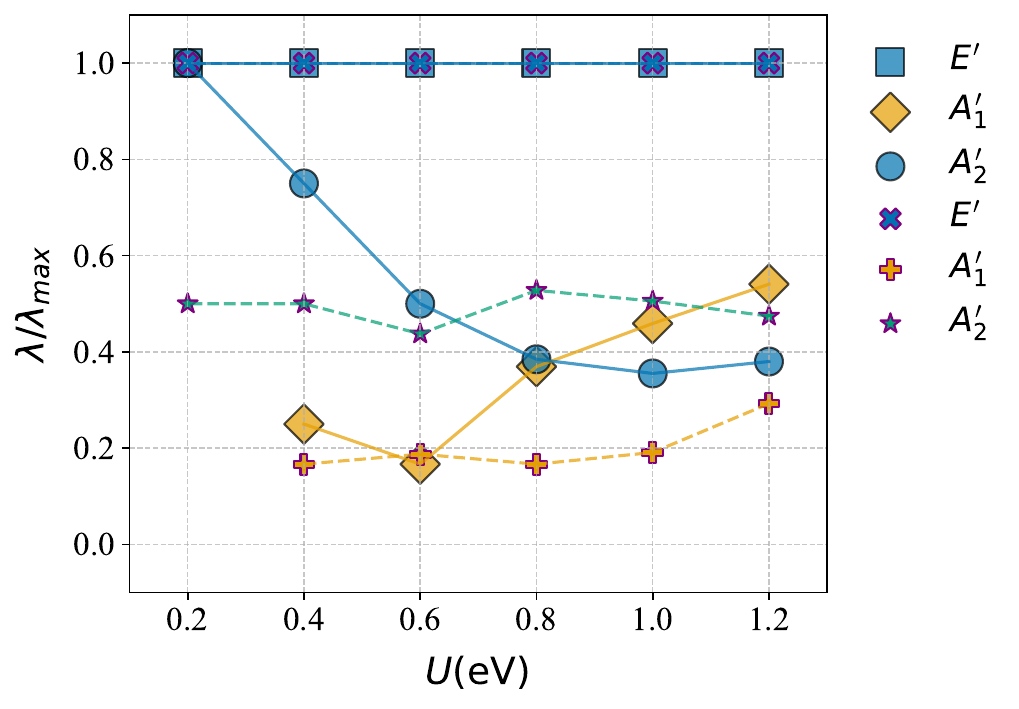}
    \caption{The evolution of the eigenvalues $\lambda/\lambda_{max}$, corresponding to specific Irreps, is presented as a function of $U$ and $J$ in the presence of Ising SOC. The four largest eigenvalues are plotted. Markers with black edge color represent $J = U/4$, while smaller markers with a purple edge color correspond to $J = U/6$.}
    \label{fig_irrep_vs_u_soc}
\end{figure}

The basis function for E${'}$ irrep is a mixture of the E$_{2g}$ and E$_{1u}$ symmetry basis functions of the D$_{6h}$ point group symmetry (symbolically also represented as $d+f$) and therefore can be considered to be a mixture of even and odd parity eigenstates (we quantify the relative mixing between even and odd parity states by the parameter $-1\leq \eta \leq 1$ with the lower (upper) bound representing a pure odd (even) parity state. See Appendix C for a detailed description of the $\eta$ parameter). As seen in Fig.~\ref{fig_gap_soc}(c), the first subdominant solution belongs to the A$_{2}'$ symmetry which is followed by solutions with E$'$ and A$_{2}'$ symmetries. As shown in Fig.~\ref{fig_irrep_vs_u_soc}, the E${'}$ symmetry solution remains the dominant instability with variation in the Coulomb interaction ($U$) and Hund's coupling ($J$).

%%%%%%%%%%%%%%%%%%
%%%%%%%%%%%
\subsubsection{Effects of electron-phonon coupling on gap solutions}
Das {\it et al.}~\cite{das2023electron} investigated the effects of both conventional electron-phonon-mediated pairing and unconventional spin-fluctuation-mediated pairing in Ising superconductors.
The authors  found that the electron-phonon coupling ($\Lambda_{}$) is highly anisotropic, with the dominant contribution arising from same-spin $K$-$K'$ scattering, leading to a constant-sign gap function. In contrast, the spin-fluctuation coupling results in a sign-changing gap function. Using first-principles calculations within the Eliashberg formalism, they computed the coupling strengths. %In this formalism, the pairing interactions remain constant between different Fermi surface pockets, lacking full momentum dependence.
Here we incorporate the electron-phonon coupling ($\Lambda_{}$) from Das {\it et al.}~\cite{das2023electron}, in conjunction with our fully momentum-dependent spin- and charge-fluctuation-mediated pairing interaction. Our goal is to investigate their combined effects on possible superconducting instabilities. We define the effective pairing vertex by
\begin{equation}
    V_{\text{Total}} = V_{e-ph} \cdot (1 - \alpha) + V_{SF} \cdot \alpha,
\end{equation}
where $\alpha$ is a tuning parameter that controls the relative mixing and runs from 0 to 1. The electron-phonon pairing interaction is given by
\begin{equation}
    V_{e-ph}(\mathbf{k},\mathbf{k}')= \frac{1}{(2\pi)^{2}} \frac{\Lambda(\mathbf{k}_{f},\mathbf{k}_{f}')}{|\nabla 
    \epsilon_{\mathbf{k}_{f}'}|} d\mathbf{k}_{f}'.
\end{equation}

\begin{figure}
\centering
    \includegraphics[width=1.0\linewidth]{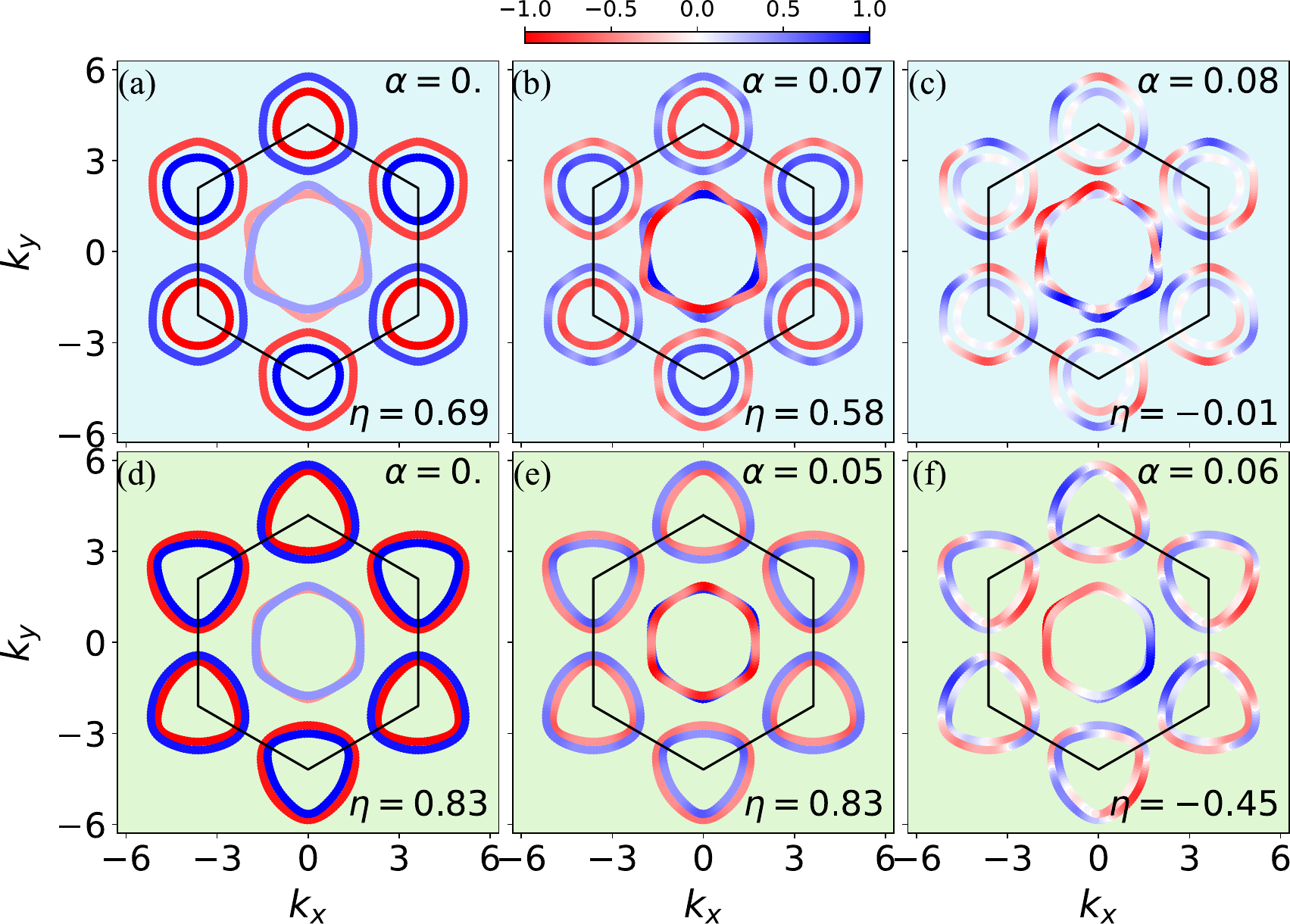}
    \caption{Superconducting instability in presence of Ising SOC as a function of $\alpha$.
    Here, $V_{SC}=(1-\alpha)V_{e-ph}+\alpha V_{SF}$. Additional parameters: $U = 0.6$ eV and $J = U/4$. The upper panels (a–c) correspond to TaS\(_2\) , while the lower panels (d–f) correspond to NbSe\(_2\), respectively.
}
    \label{fig_e_ph_with_soc}
 \end{figure}

As shown in Fig.~\ref{fig_e_ph_with_soc}, the sign change in the gap structure satisfies the condition of the exclusion principle 
($\Delta^{1,2}_{\uparrow \downarrow}(\vk)=-\Delta^{2,1}_{\downarrow \uparrow}(-\vk)$). As shown in Fig.~\ref{fig_e_ph_with_soc}(a,d), for pure electron phonon interaction, the gap function magnitude over each Fermi pocket is quite isotropic, but has an appreciable odd parity mixing that increases with spin-orbit interaction strength ($\eta$ is smaller in TaS$_2$ compared to NbSe$_2$). With increase of $\alpha$ [Figs.~\ref{fig_e_ph_with_soc}(b), \ref{fig_e_ph_with_soc}(c), \ref{fig_e_ph_with_soc}(e), \ref{fig_e_ph_with_soc}(f)], the dominant gap function shifts to the E$^{'}$ symmetry solution for $\alpha \ge 0.07$ for TaS\(_2\) and $\alpha \ge 0.05$ for NbSe\(_2\). If we consider that the superconducting pairing interactions can be decomposed into a sum of contributions from each Irrep in general, we can expect the e-ph and spin-fluctuation-mediated pairing channels to add up according to their contribution to each symmetry channel. Since the e-ph pairing contribution is attractive and constant [see Appendix D], it would add to the repulsive subdominant constant contribution from the spin-fluctuation pairing. This can explain the rapid transition of the dominant gap solution from a A$_1^{'}$ symmetry to E${'}$ symmetry solution. Having obtained the detailed superconducting gap structure, we now test the significance of these results by comparing them to recent experiments on monolayer TaS$_2$.
\begin{figure*}
\centering
    \includegraphics[width=1.0\linewidth]{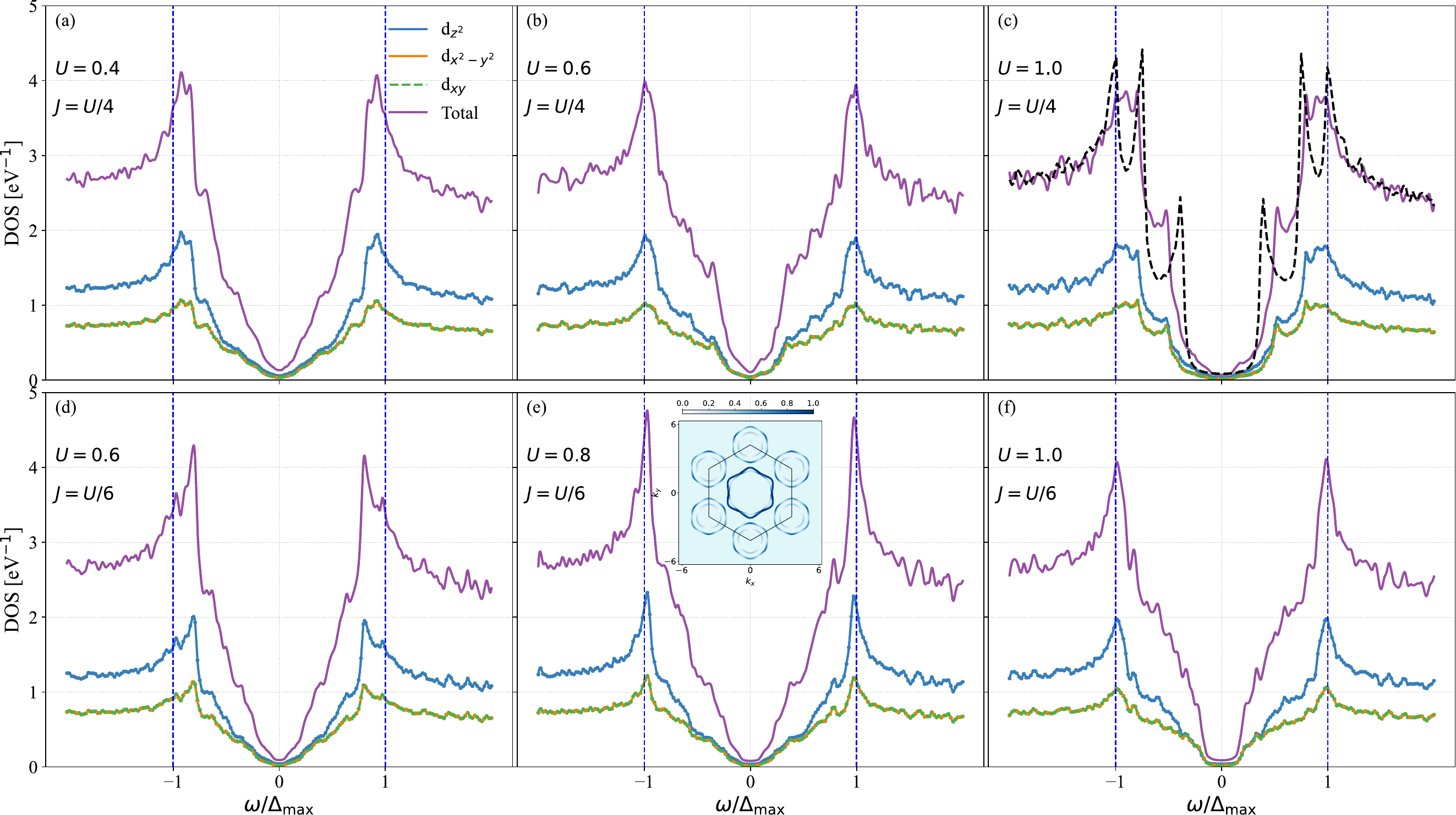}
    \caption{(a) Calculated orbital resolved density of states for various values of $U$. (a), (b), and (c) correspond to $J=U/4$, whereas (d), (e), and (f) are for $J=U/6$. In panel (c) the DOS for a pure electron phonon driven pairing interaction is also presented as a black dashed line. The DOS is predominantly characterized by contributions from the \( d_{z^{2}} \) orbital.
}
    \label{fig_dos1}
 \end{figure*}
 
The first measurable quantity analyzed in this work is the superconducting density of states (DOS) [see Appendix D for details], which can be directly compared to low temperature STM experiments that find a nodal local DOS in superconducting TaS$_2$~\cite{Vano2023,Cho2022}. However, in order to maximize the condensation energy, the superconducting ground state for the two dimensional irrep E${'}$ is likely to choose the $\Delta(\vk)=\Delta_1(\vk) + i\Delta_2(\vk)$ solution, where $\Delta_1(\vk)$ and $\Delta_2(\vk)$ are the degenerate superconducting ground state solutions of the linearized gap equation shown in Fig.~\ref{fig_gap_soc}(a,b). 

\begin{figure}
\centering
    \includegraphics[width=1.0\linewidth]{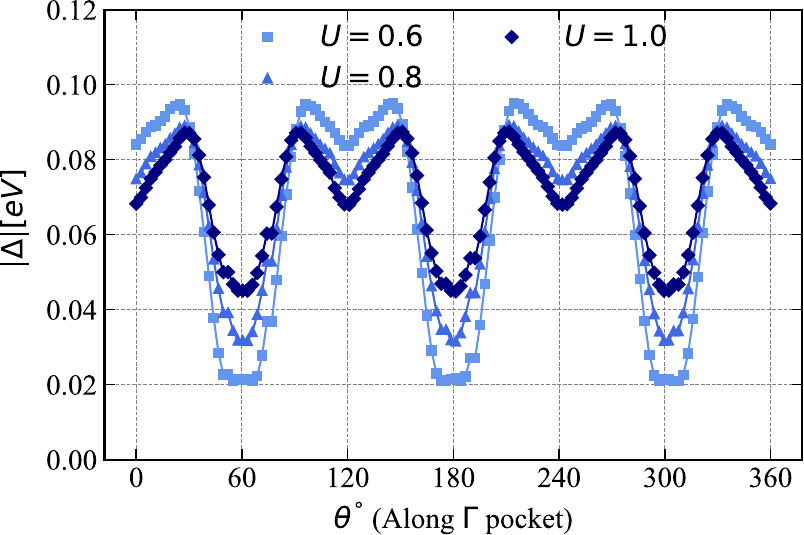}
    \caption{Superconducting gap magnitude in \((1,i)\) state plotted along the spin up Fermi surface for the $\Gamma$-centered pocket as a function of polar angle $\theta$ for different $U$ values, and \(J=U/4\). As can be seen the gap is more anisotropic for lower values of \(U\).
}
    \label{fig_gapaniso}
 \end{figure}

As shown in Fig.~\ref{fig_dos1}(e) inset, the magnitude of this $(1,i)$ state belongs to the A$_{1}'$ representation of the D$_{3h}$ point group symmetry, as expected from the decomposition of the reducible representation $\Gamma_{E'} \times \Gamma_{E'}^*$. In Fig.~\ref{fig_dos1} the DOS is shown for a range of Coulomb interaction strengths $U$ for $J=U/4$ in the Figs.~\ref{fig_dos1}(a)-\ref{fig_dos1}(c) and $J=U/6$ in Figs.~\ref{fig_dos1}(d)-\ref{fig_dos1}(f). We find that with decreasing $U$, the DOS becomes more nodal. The features in the DOS agrees with experimental observations for $U\leq 0.6$ eV for both $J=U/4$ and $J=U/6$~\cite{Vano2023,Cho2022}. The variation in the DOS with Coulomb interaction can be understood from Fig.~\ref{fig_gapaniso}, where we show that although the gap magnitude of the $(1,i)$ state does not have explicit nodes, the gap anisotropy over the Fermi surface monotonically increases with decreasing $U$. This behavior can be compared with the DOS for a purely electron-phonon-driven superconductivity, where, as expected, we find a fully opened gap, as shown in Fig.~\ref{fig_dos1}(c) [black dashed lines]. Therefore, in summary the addition of spin-fluctuation-mediated pairing transforms a conventional full-gap structure into an unconventional nodal-like form in agreement with experiments~\cite{Vano2023,Cho2022}.

%\subsubsection{Even-Odd Parity Mixing}
Another observation that has generated significant interest is the observation of Leggett modes in STM experiments on monolayer NbSe$_2$~\cite{wan2022observation} and \ts{}~\cite{Vano2023}. It has been argued that the presence of the mode is a consequence of  even-odd parity mixing~\cite{bittner2015leggett} that is allowed in superconductors with broken inversion symmetry, or originates from multigap physics due to differences in gap magnitudes between the $\Gamma-$ and $K-$centered Fermi pockets~\cite{das2023electron}. Here, we do not find significant differences in gap magnitude between the two Fermi pockets in the presence of SOC, although the gap function does have significant mixing between even and odd-parity superconducting states. 
\begin{figure}[tb]
\centering
    \includegraphics[width=1.0\linewidth]{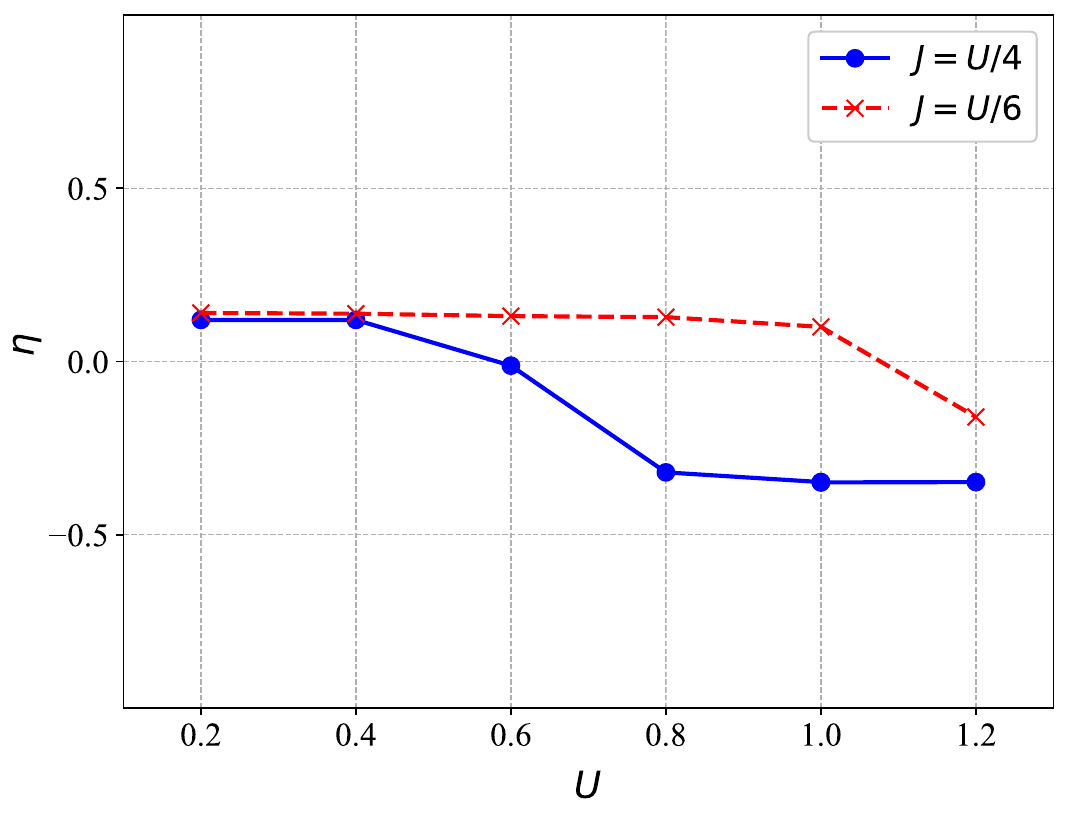}
    \caption{Even-odd parity mixing ratio $\eta$ as a function of the Coulomb interaction $U$ with $J = U/4$ and $J = U/6$ respectively.  $\eta = \pm 1$ corresponds to the pure even (odd) parity solutions.}
    \label{fig_eta}
 \end{figure}
In Fig.~\ref{fig_eta}, the variation of the mixing ratio \(\eta\) of the pure spin-fluctuations driven ground state gap function reveals that the odd-parity superconducting state dominates the mixed state for larger Coulomb interactions ($U\geq 0.6$ eV). Furthermore, compared to \nb{}~\cite{roy2024unconventional}, we find that in the case of \ts{}, the odd-parity mixing is more prominent even at lower Coulomb interaction strengths.  Our results therefore indicate that monolayer di-chalcogenide superconductors like TaS$_2$ and NbSe$_2$ have a significant admixture of even and odd parity states in the superconducting gap structure.  In the next section, we show that the \(\eta\) parameter qualitatively agrees well with studies on the magnetic field dependence.

\subsection{Magnetic field effect}
The primary effect of an applied magnetic field arises from its coupling to electron spins through the Zeeman interaction and the coupling to orbital motion of electrons. In two-dimensional materials, due to the confinement of electronic motion within the plane, the coupling of an in-plane magnetic field to orbital motion can often be neglected. 

In Ising superconductors, strong SOC enforces the spin quantization axis along the out-of-plane (\( z \)) direction, effectively pinning electron spins along this axis. As a result, the effect of Zeeman interaction from an in-plane magnetic field is significantly weakened, leading to an enhancement of the upper critical field in comparison to the Pauli limit~\cite{lu2015evidence,xi2016ising,lu2018full,de2018tuning}.

For an in-plane magnetic field (\(\mathbf{B} \perp z\)) along the $x$-direction, the Zeeman contribution can be expressed as
\begin{align}
H_{B_x}=g\mu_B B_x\sum_{\vk,l} (c^{\dag}_{\vk l \uparrow}c_{\vk l \downarrow}+c^{\dag}_{\vk l\downarrow}c_{\vk l\uparrow}).
\end{align}

A number of recent works have studied the effect of transverse magnetic fields in Ising superconductors assuming the explicit presence of spin-singlet and spin-triplet states that may have some relative mixing. Computationally, we do not make this assumption, but solve the gap equation to extract the critical magnetic field for each temperature point. The linearized gap equation in the presence of a finite in-plane magnetic field becomes\cite{roy2024unconventional}
\begin{widetext}
 \begin{align}\label{fullgapeqn}
-\Delta^{\tilde{m},\tilde{m}'}_{}(\vk)=&\sum_{\substack{\vk' }}\left[ \left(  V^{\tilde{m}',\tilde{m}}_{\tilde{m},\tilde{m}'}(\vk,\vk') I^{}_{1}(\vk')
+
   V^{\tilde{m}',\tilde{m}}_{\tilde{m}',\tilde{m}}(\vk,\vk') I^{}_{2}(\vk')\right)\Delta^{\tilde{m},\tilde{m}'}_{}(\vk')
   \right.\nonumber\\
 &+ \left.\left( V^{\tilde{m}',\tilde{m}}_{\tilde{m}',\tilde{m}}(\vk,\vk')  I^{}_{1}(\vk')
  + V^{\tilde{m}',\tilde{m}}_{\tilde{m},\tilde{m}'}(\vk,\vk') I^{}_{2}(\vk')\right)\Delta^{\tilde{m}',\tilde{m}}_{}(\vk')\right],
\end{align}
where
\begin{subequations}
\begin{align}
    I_{1}(\vk)&=- \text{ln}\left(\frac{T_{}}{T_{\mathrm{c_{}}}}\right)+\frac{1}{\lambda}
%     \nonumber\\&
    - \frac{1}{2}\left(\frac{-h_{\vk}^{2}}{h_{\vk}^{2}+\delta_{\vk}^{2}}\right)
    \left\{\mathrm{Re}[\psi(\frac{1}{2}+i\frac{\sqrt{h_{\vk}^{2}+\delta_{\vk}^{2}}}{2\pi k_{B}T_{}})] -\psi(\frac{1}{2})\right\},\\
      I_{2}(\vk)&=\frac{1}{2}\left(\frac{-h_{\vk}^{2}}{h_{\vk}^{2}+\delta_{\vk}^{2}}\right)\left\{\mathrm{Re}[\psi(\frac{1}{2}+i\frac{\sqrt{h_{\vk}^{2}+\delta_{\vk}^{2}}}{2\pi k_{B}T_{}})] -\psi(\frac{1}{2})\right\}.\label{eq_I12}
\end{align}
\end{subequations}
\end{widetext}
Here, $\psi$ denotes the digamma function, $2\delta_{\vk}$ represents a momentum-dependent energy gap between two SOC split bands near the Fermi level, $\lambda$ is the eigenvalue of the dominant gap equation solution in zero magnetic field, and $T_c=3$K. Further detailed steps can be found in the Appendix.

In the presence of Ising SOC, the results become non-trivial due to the mixing of momentum-dependent even- and odd-parity superconducting states, as well as the influence of a momentum-dependent SOC. 
Although the transition temperature is modulated by the magnetic field, the dominant gap function remains unchanged from the zero-field gap structure (except for a splitting of the degenerate E' Irrep solution discussed later). 

\begin{figure}
\centering
    \includegraphics[width=1.0\linewidth]{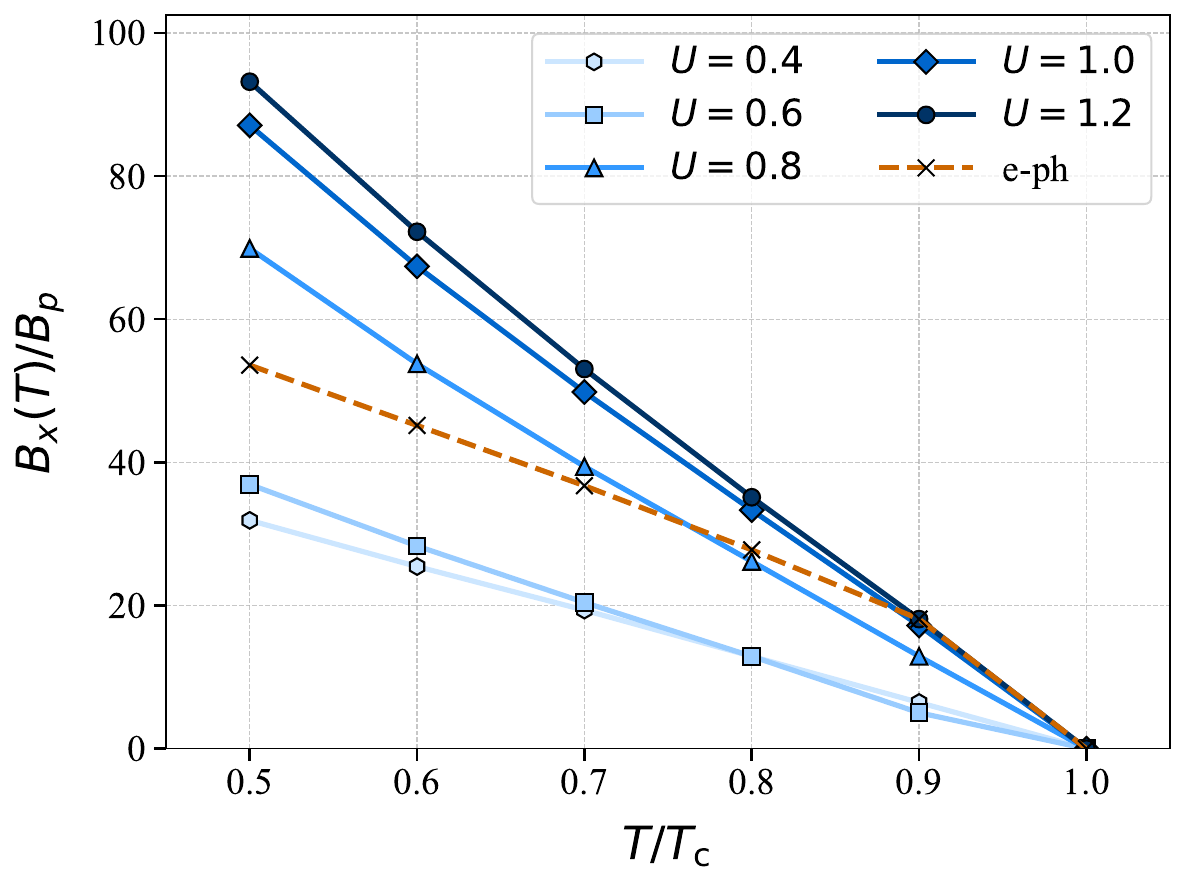}
    \caption{Effect of an in-plane magnetic field on $T/T_{c}$ in the presence of SOC for various $U$ in a spin-fluctuation-driven pairing scenario. The dashed line corresponds to the pure electron-phonon mechanism. Here, $B_{p}$ is the Chandrasekhar-Clogston (or Pauli) paramagnetic limit and $B_{p}=(1.86 T K^{-1})T_{c}$.
}
    \label{fig_magnetic_so_upc}
 \end{figure}

In Fig.~\ref{fig_magnetic_so_upc}, we present the critical magnetic field curves for various superconducting pairing interactions. The curves show a significant enhancement of the critical magnetic field in comparison to the Pauli limit. Additionally, we also find the critical magnetic field increases with \( U \), whereas for a pure electron-phonon pairing it lies in the intermediate region. This behavior aligns well with the values of the mixing parameter \( \eta \) [see Fig.~\ref{fig_eta}], where we find that as \( U \) increases, the odd-parity component becomes more prominent. These results qualitatively agree with the observed enhancement in upper critical field in TaS$_2$ superconductor. We note at this point that the perturbative approach as outlined in Appendix \ref{app_B} is only accurate for small fields and superconductivity at large fields will be suppressed if pairing is recalculated from the susceptibility in the presence of a magnetic field.
%%%%%%%
\begin{figure}
\centering
    \includegraphics[width=1.0\linewidth]{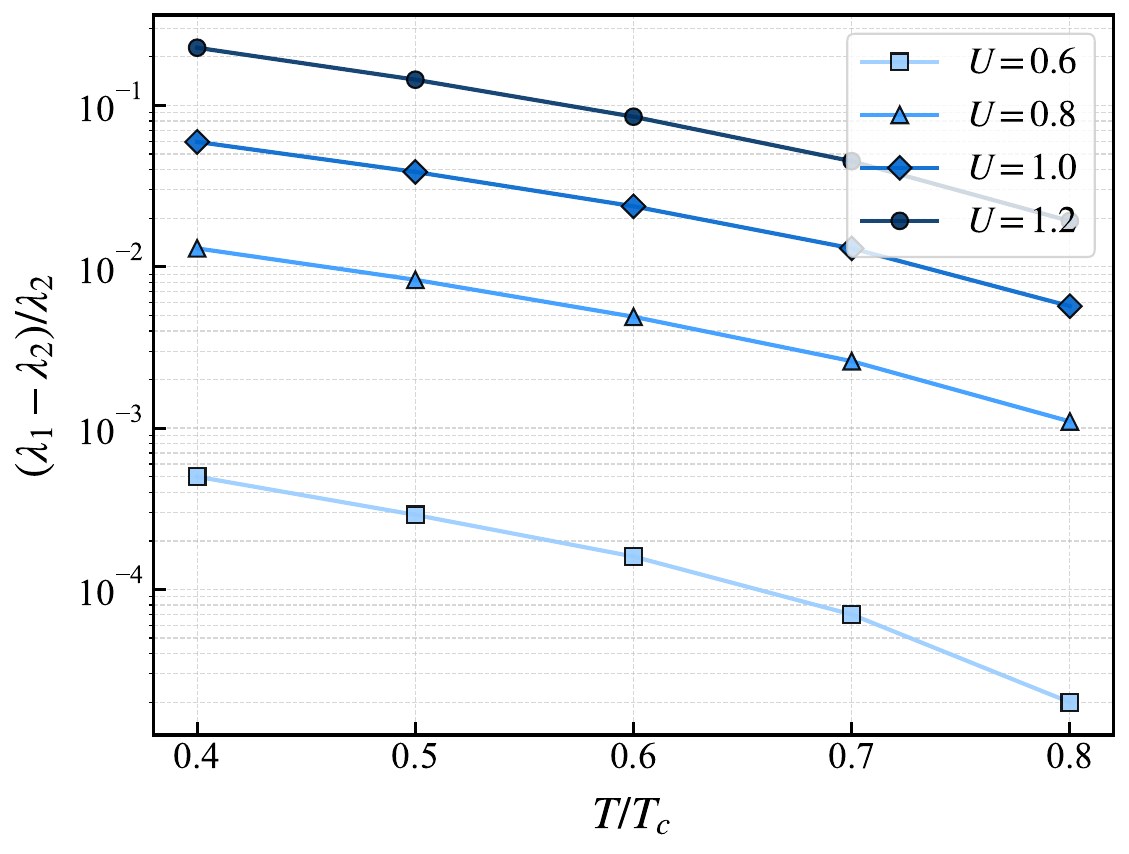}
    \caption{The lifting of the degeneracy of the two-dimensional E\('\) irreducible representation in the presence of an in-plane magnetic field. The critical value of the in-plane magnetic field for a given \(T/T_c\) can be obtained from Fig.~11. 
    %Inset shows the absence of splitting for the same calculation, but in the presence of an odd parity degenerate solution and absence of Ising SOC.% \AK{Choice of symbols/colors is similar to Fig. 11, but does not match. Can we have same dataset (U-values) in Fig. 11 and 12?}\SR{For U=0.4 the splitting is of the ordrer of 10$^{-6}$}
}
    \label{fig_splitting}
 \end{figure}
%%%%%%%

Another experiment of interest that hints towards the structure of ground state superconducting gap in monolayer TMD superconductors has been the observation of twofold symmetry in magnetoresistance measurements with in-plane rotated magnetic field~\cite{hamill2021two,Cho2022}. It was proposed that the twofold symmetry could either be explained by a ground state belonging to the two representation E${'}$ or E${''}$ Irrep of the D$_{3h}$ point group, or through coupling of an A$_1^{'}$ representation to the multidimensional representation through some symmetry breaking field like strain from a substrate layer~\cite{hamill2021two}.
 
\begin{figure}[htbp]
\centering
    \includegraphics[width=1.0\linewidth]{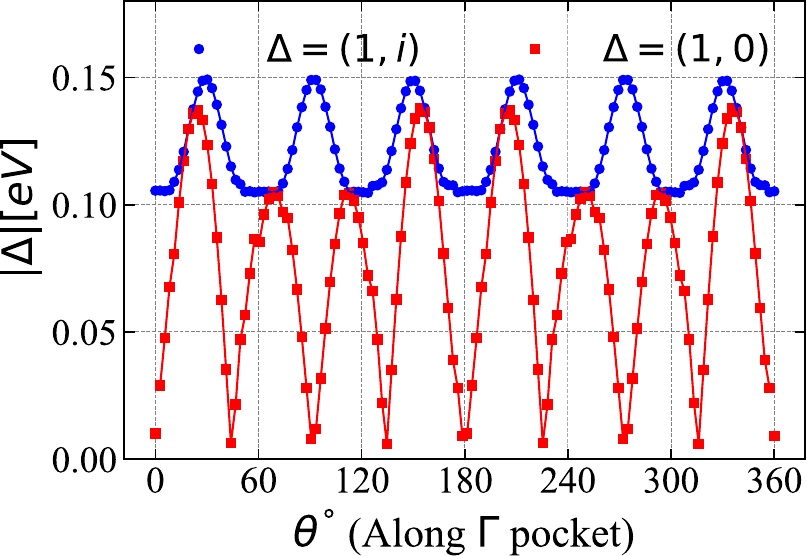}
    \caption{The absolute values of the gap functions corresponding to the \((1,i)\) and \((1,0)\) states are shown, respectively. The \((1,0)\) state can be associated with the twofold symmetry observed in magnetoresistance experiments. 
}
    \label{fig_2fold}
 \end{figure}

 Although, the dominant superconducting instability obtained in our calculations belongs to the E$^{'}$ Irrep, the magnitude of the expected $(1,i)$ ground state solution will not break the lattice symmetry [see Fig.~\ref{fig_dos1}(e) inset]. However, as shown in Fig.~\ref{fig_splitting}, interestingly the eigenvalues of the two degenerate solutions $(\lambda_1,\lambda_2)$ split in a finite transverse magnetic field, and the splitting is larger for larger $U$ in general. In contrast, no such splitting occurs in the absence of
 Ising SOC, where even-odd parity mixing is forbidden. This implies that the dominant instability would shift from the sixfold symmetric $(1,i)$ state to the twofold symmetric $(1,0)$ state in finite fields. We show the angular dependence of the $(1,i)$ state and $(1,0)$ state over the $\Gamma $ centered Fermi pockets in Fig.~\ref{fig_2fold}. This result therefore reveals a direct mechanism for the observation of twofold symmetry in the superconducting gap structure in transverse fields in monolayer NbSe$_2$ and monolayer TaS$_2$. 

%%%%%%%%%%%%%%%%%%%%%%%%%%%%%%%%%%%%%%%%%%%%%%%%%%%%%%%%%%%%%%%
\section{Summary} \label{sec:level4}
In summary, motivated by experimental evidence for unconventional pairing in monolayer dichalcogenide superconductors, we have calculated the dominant superconducting instability in monolayer Ising superconductors like TaS$_2$ and NbSe$_2$, assuming a momentum-dependent spin and charge fluctuation mediated pairing mechanism. The calculations consider realistic low-energy tight-binding Hamiltonians extracted from DFT, and incorporate electronic correlations from a Hubbard-Hund Hamiltonian. Solving the linearized superconducting gap equation, we find that the superconducting ground state belongs to a two-dimensional E$^{'}$ Irrep. We explore the parameter space by studying the effects of changes in pairing interaction due to variations in Coulomb interaction and find the dominant E$^{'}$ symmetry ground state superconducting solution remains robust for the TaS$_2$ Hamiltonian.

We have further modeled physical quantities relevant to a number of recent experimental observations and found the following. 1) The superconducting density of states features a node-like form similar to observations in STM experiments ~\cite{Vano2023,Cho2022}. 2) A strong enhancement of in-plane upper critical field compared to Pauli limit seen in experiments studying magnetic field dependence~\cite{lu2015evidence,xi2016ising,lu2018full}. 3) A significant mixing of even and odd parity superconducting states. The odd-parity contribution increases with $U$, and these results could provide an explanation for the observation of Leggett modes in tunneling experiments, and 4) with a finite in-plane magnetic field, the dominant superconducting instability transforms from a $(1,i)$ to a $(1,0)$ solution of the two-dimensional $E'$ irrep. The $(1,0)$ solution has a twofold symmetric gap structure that can provide a natural explanation for the observed anisotropy seen in magnetoresistance measurements~ \cite{hamill2021two,Cho2022,liu2024nematic}. Although alternative theoretical scenarios have been proposed previously to explain individual experiments on the monolayer Ising superconductors, by collectively providing an explanation for disparate experimental observations, we believe this work provides a promising theoretical scenario for the superconducting state in Ising superconductors realized in monolayer NbSe$_2$ and monolayer TaS$_2$.

\begin{acknowledgments}
A.K.~acknowledges support by the Danish National Committee for Research Infrastructure (NUFI) through the ESS-Lighthouse Q-MAT. A.K.~acknowledges support by the Institute of Eminence (IoE) program of IIT Madras. S. M. and S. R. acknowledge support from IIT Madras through HRHR travel mobility grant. B.M.A. acknowledges support from the Independent Research Fund Denmark Grant No. 5241-00007B and a research grant (VIL69220) from VILLUM FONDEN.
\end{acknowledgments}

\appendix

%\section{Introduction}
\section{Band structure and Fermi surface}
\label{app_A}
%%%%%%%%%
\begin{figure}[h]
    \includegraphics[width=1.0\linewidth]{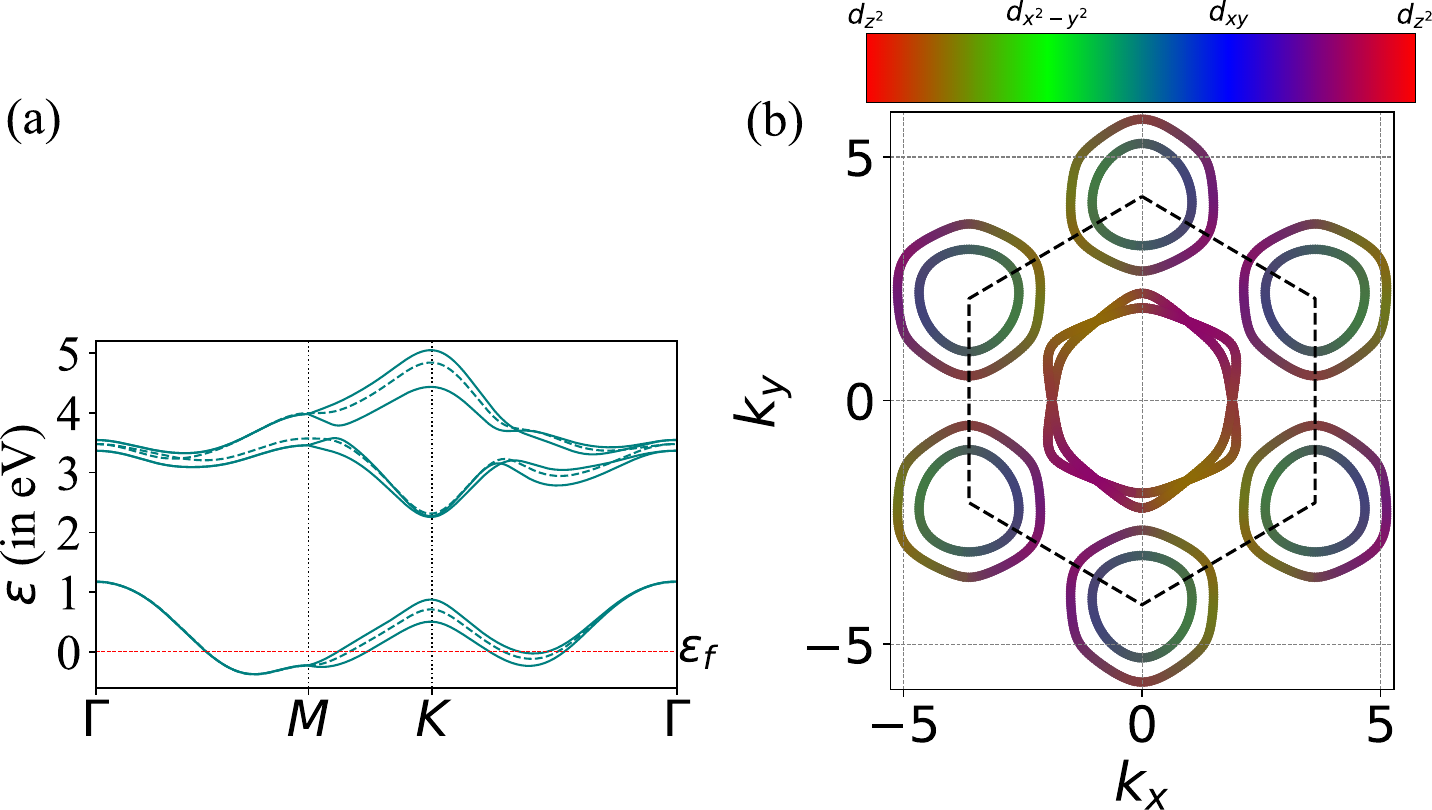}     
    \caption{(a) The band structure with full bandwidth. (b) The Fermi surface colored according to the orbital content.}
    \label{fig_band_fs}
\end{figure}
To obtain the electronic structure of a monolayer TaS$_2$, we set up a DFT calculation using the full-potential local-orbital (FPLO) code \cite{Koepernik1999}, version 22.00-62 where a monolayer TaS$_2$ was set up with the lattice constants $a=b=3.337\mathrm{\AA{}}$ and $c=23.15\mathrm{\AA{}}$ to simulate a vacuum. The space group is \# 187, i.e. P-6m2 with the Ta at the 1b Wyckoff postion $(0,0,1/2)$ and the Se atom at 2h $(1/3,-1/3,0.433)$ as obtained from a optimization procedure to minimize the energy. We employed the scalar relavistic setting for the band structure without SOC and the fully relativistic setting to obtain the effects of Ising SOC, both with the GGA functional. The Wannier projection was performed onto the three orbitals $d_{z^2}$, $d_{x^2-y^2}$, $d_{xy}$ with energy window of $[-0.4,3.5]\,\mathrm{eV} \pm 0.5 \mathrm{eV}$. The numerical values of the resulting tight-binding parameters are available in Ref. \cite{tb_git}.

Figure~\ref{fig_band_fs}(a) shows Ising SOC spilt bands at obtained from the corresponding tight-binding model. There are a total six bands, where the lower two are half-filled. We primarily focus on low energy bands crossing the Fermi surface. Figure~\ref{fig_band_fs}(b) shows orbital resolved Fermi surface. As can be seen in figure $\Gamma$-centered pocket is primarily dominated by the $d_{z^{2}}$ orbital and the K-centered pockets are primarily dominated by degenerate $d_{x^{2}-y^{2}}$ and $d_{xy}$ orbitals.

\section{Effective pairing interaction}
\label{app_B}
The interaction Hamiltonian is given by
\begin{eqnarray}
H_{\mathrm{int}}&=&\frac{U}{2} \sum_{i,\mu,\sigma} n_{i \mu \sigma}n_{i \mu \Bar{\sigma}}
+\frac{U'}{2} \sum_{i,\mu \neq \nu , \sigma} n_{i \mu \sigma }n_{i \nu \Bar{\sigma}}\nonumber\\
&+&\frac{U'-J}{2} \sum_{i,\mu \neq \nu , \sigma} n_{i \mu \sigma }n_{i \nu \sigma} \nonumber\\
&+&\frac{J}{2} \sum_{i,\mu \neq \nu}   \sum_{\sigma} c^{+}_{i \mu \sigma} c^{+}_{i \nu \Bar{\sigma}} c^{}_{i \mu \Bar{\sigma}} c^{}_{i \nu \sigma}   \nonumber\\
&+&\frac{J'}{2} \sum_{i,\mu \neq \nu}   \sum_{\sigma}    c^{+}_{i \mu \sigma} c^{+}_{i \mu \Bar{\sigma}} c^{}_{i \nu \Bar{\sigma}} c^{}_{i \nu \sigma}
\end{eqnarray}
Here, the $U$, $U'$, and $J$ terms denote the intraorbital, interorbital Hubbard
repulsion and the Hund’s rule coupling as well as the pair hopping. 
We work in the spin-rotational invariant setting where the onsite Coulomb terms are related by $U = U' + 2J$ and $J=J'$.
Restricting to the Cooper pair channel the above interaction Hamiltonian can be rewritten with the compact notation  $[\tilde{l}:=(l,\sigma)]$ as following
\begin{align}
H_{\mathrm{int}}=\frac{1}{2} \sum_{\vk,\vk',\tilde{l}} [\hat{U}]^{\tilde{l}_1,\tilde{l}_2}_{\tilde{l}_3,\tilde{l}_4} \, c^{\dagger}_{\vk,\tilde{l}_1} \,c^{\dagger}_{-\vk,\tilde{l}_3} \, c_{-\vk',\tilde{l}_2}\, c_{\vk',\tilde{l}_4}
\label{eqn:int_hamil_u}
\end{align}
The bare electron-electron interaction $ [\hat{U}]^{\tilde{l}_1,\tilde{l}_2}_{\tilde{l}_3,\tilde{l}_4}$, i.e. the first order correction in $ [\hat{V}]^{\tilde{l}_1,\tilde{l}_2}_{\tilde{l}_3,\tilde{l}_4}$ is given by
\begin{center}
\begin{tabular}{c c c}
$[\hat{U}]_{\mu \Bar{s} \mu s}^{\mu s \mu \Bar{s}} = U$ & $[\hat{U}]_{\mu \Bar{s} \nu s}^{\nu s \mu \Bar{s}}= U'$ & $[\hat{U}]_{\mu \Bar{s} \nu s}^{\mu s \nu \Bar{s}}= J'$ \\
                                                &                                                                                                     \\
$[\hat{U}]_{\nu \Bar{s} \nu s}^{\mu s \mu \Bar{s}} = J$ &\,\,\,\,\,\,\,\,\,\, $[\hat{U}]_{\nu s \mu s}^{\mu s \nu s}= U'-J$           & \\
\\
 \,\,\,\,\,\,$[\hat{U}]_{\mu \Bar{s} \mu \Bar{s}}^{\mu s \mu s} =-U $ &\,\,\,\,\,\,$[\hat{U}]_{\mu \Bar{s} \mu \Bar{s}}^{\nu s \nu s}=-U'   $ &\,\,\, $[\hat{U}]_{\mu \Bar{s} \nu \Bar{s}}^{\mu s \nu s}= -J'$\\ 
                                                                                                            \\
\,\,\,\,\,\,$[\hat{U}]_{\nu \Bar{s} \mu \Bar{s}}^{\mu s \nu s} =-J $ &\,\,\,\,\,\,\,\,\,\,\,\,\,\,\,\,\,$[\hat{U}]_{\nu s \nu s}^{\mu s \mu s}=-U'+J $ &  
\end{tabular}
\end{center}

%%%%%%%%%
For higher order correction in $ \hat{V}$ we sum up all bubble and ladder diagrams to infinite order in $\hat{U}$ (RPA).

In presence of spin-orbit coupling the final interaction vertex can be written as~\cite{roy2024unconventional}
% \begin{widetext}

\begin{align}
[\hat{V}(\vk,\vk')]^{\tilde{l}_1,\tilde{l}_2}_{\tilde{l}_3,\tilde{l}_4}
  &= [\hat{U}]^{\tilde{l}_1,\tilde{l}_2}_{\tilde{l}_3,\tilde{l}_4}
   - \bigg[\frac{\hat{U}\,\hat{\chi_{0}}\,\hat{U}}
                 {1-\hat{\chi_{0}}\,\hat{U}}
     \bigg]^{\tilde{l}_1,\tilde{l}_4}_{\tilde{l}_3,\tilde{l}_2}
     (\vk-\vk') \nonumber \\
  &\quad
   + \bigg[\frac{\hat{U}\,\hat{\chi_{0}}\,\hat{U}}
                 {1-\hat{\chi_{0}}\,\hat{U}}
     \bigg]^{\tilde{l}_1,\tilde{l}_2}_{\tilde{l}_3,\tilde{l}_4}
     (\vk+\vk') .
\label{eqn:final_vertex}
\end{align}

This is the pairing vertex used in Eq.~(3) of the main text in the
superconducting Hamiltonian,
\begin{align}
H_{\rm int}
   &= \tfrac{1}{2} \sum_{\vk,\vk',\tilde{l}}
   [\hat{V}]^{\tilde{l}_1,\tilde{l}_2}_{\tilde{l}_3,\tilde{l}_4}
   (\vk,\vk') \,
   c^{\dagger}_{\vk,\tilde{l}_1} \,
   c^{\dagger}_{-\vk,\tilde{l}_3} \,
   c_{-\vk',\tilde{l}_2}\,
   c_{\vk',\tilde{l}_4}.
\label{eqn:int_hamil_b}
\end{align}

\begin{widetext}
%%%%%%%%%%%%%%%%%%%%%%%%%%%%%%%%%%
%%%%%%%%%%%%%%%%%%%%%%%%%%%%%%%%%%
\section{Gap equation for multiband superconductors}

%Assuming a homogeneous superconducting state,
% \AK{This assumption goes in later!}\SR{Do you mean this should be starting line of Appendix B?}
The Hamiltonian for a multi band superconductor can be expressed as,
\begin{eqnarray}
H&=&H_0+H_{\mathrm{sc}}\\
H_0&=&\sum_{\vk n \sigma}\epsilon_{n\sigma}(\vk)a^{\dag}_{\vk n \sigma}a_{\vk n \sigma}\label{eqn:h0}
\end{eqnarray}
% \begin{widetext}
\begin{eqnarray}
H_{\mathrm{sc}}&=&\frac{1}{2}\sum_{n_i\sigma_i \vk \vk'} V^{n_1,n_2,\sigma_1,\sigma_2}_{n_3,n_4,\sigma_3,\sigma_4}(\vk,\vk')a^{\dag}_{-\vk n_1 \sigma_1}a^{\dag}_{\vk n_2 \sigma_2}a_{\vk' n_3 \sigma_3}a_{-\vk' n_4 \sigma_4}\label{eqn:hsc}
\end{eqnarray}
Here,
\begin{eqnarray}
   V^{n_1,n_2,\sigma_1,\sigma_2}_{n_3,n_4,\sigma_3,\sigma_4}(\vk,\vk')=\sum_{\{l_1,l_2,l_3,l_4\}} u^{n_{1}*}_{l_{1},\sigma_{1}}(-\vk)u^{n_{3}*}_{l_{3},\sigma_{3}}(\vk)V^{l_1,l_2,\sigma_1,\sigma_2}_{l_3,l_4,\sigma_3,\sigma_4}(\vk,\vk')u^{n_{2}}_{l_{2},\sigma_{2}}(\vk')u^{n_{4}}_{l_{4},\sigma_{4}}(-\vk')
\end{eqnarray}
\end{widetext}
where the pairing vertex on the left hand side of the above equation is in the band basis whereas it is expressed in the orbital basis on the right hand side. For ease of notation we have changed the convention of writing down the pair potential in orbital basis, $ [\hat{V}]^{\tilde{l}_1,\tilde{l}_2}_{\tilde{l}_3,\tilde{l}_4}$ in Eq. \ref{eqn:hsc} is the same as $ [\hat{V}]^{\tilde{l}_2,\tilde{l}_4}_{\tilde{l}_1,\tilde{l}_3}$ in Eq. \ref{eqn:final_vertex}. Note that in the following the combined notation $\tilde{l}_i=(l_i,\sigma_i)$.  

The above Hamiltonian is written in a multiband basis where $a_{\vk n \sigma}$($a_{\vk n \sigma}^\dagger$) annihilates (creates) an electron in band $n$, at momenta $\vk$, and quantum number $\sigma$. If the non interacting Hamiltonian preserves the $S_z$ quantum number then we can identify $\sigma$ with the true spin, $\sigma=(\uparrow,\downarrow)$. This would be true for example either in the absence of a spin-orbit coupling or presence of a spin-orbit coupling that does not generate off diagonal matrix elements (spin flip contributions). In the presence of spin flip contributions $\sigma$ would represent a pseudo spin index.

Assuming a homogeneous superconducting state, the linearized gap equation for a system with Ising SOC system can be written as~\cite{roy2024unconventional}
\begin{widetext}
\begin{eqnarray}
    \begin{pmatrix}
 \Delta^{1,2}_{\uparrow,\downarrow}(\vk_{F_{1}}) \\
 \Delta^{2,1}_{\downarrow,\uparrow}(\vk_{F_{2}})
\end{pmatrix}
=-\ln\left(\frac{2\omega_c e^{\gamma}}{\pi k_{\mathrm{B}}T_c}\right)
\begin{pmatrix}
 \frac{V^{2,1,\downarrow,\uparrow}_{1,2,\uparrow,\downarrow}(\vk_{F_{1}},\vk_{F_{1}}')}{|\nabla\epsilon_{1,\uparrow}|_{\vk'=\vk_{F_{1}}'}} & \frac{V^{2,1,\downarrow,\uparrow}_{2,1,\downarrow,\uparrow}(\vk_{F_{1}},\vk_{F_{2}}')}{|\nabla\epsilon_{2,\downarrow}|_{\vk'=\vk_{F_{2}}'}} \\
 \frac{V^{1,2,\uparrow,\downarrow}_{1,2,\uparrow,\downarrow}(\vk_{F_{2}},\vk_{F_{1}}')}{|\nabla\epsilon_{1,\uparrow}|_{\vk'=\vk_{F_{1}}'}} & \frac{V^{1,2,\uparrow,\downarrow}_{2,1,\downarrow,\uparrow}(\vk_{F_{2}},\vk_{F_{2}}')}{|\nabla\epsilon_{2,\downarrow}|_{\vk'=\vk_{F_{2}}'}} 
\end{pmatrix}
   \begin{pmatrix}
 \Delta^{1,2}_{\uparrow,\downarrow}(\vk_{F_{1}}') \\
 \Delta^{2,1}_{\downarrow,\uparrow}(\vk_{F_{2}}'),
\end{pmatrix}
\end{eqnarray}
\end{widetext}
where $\omega_c$ is some cutoff that we assume to be the same for both band integrals. Note that for N $\vk_{F_{1}}$ points, and N $\vk_{F_{2}}$ points, the dimension of the gap equation matrix is $2N \times 2N$\footnote{This form of the linearized gap equation (in absence of spin-orbit coupling) has been discussed widely for example in Refs.~\cite{Graser2009,wang2013superconducting,kemper2010sensitivity,wu2015} where the Fermi surface integral is discretized as well. However, this last step of writing the equation as matrix equation is often not given explicitly.}.  The above linearized superconducting gap equation can be solved to determine the ground state gap solution. The generic gap solution with spin-orbit coupling mixes the even and odd parity states.

In order to quantify the relative mixing between even- and odd-parity contributions, we first note that in Ising superconductors, the eigenstates cannot be explicitly decomposed into even and odd parity contributions, since although a gap $\Delta_{\uparrow \downarrow}^{1,2}(\vk)$ is well defined, we cannot define the pair state with just the spin index exchanged due to spin-momentum locking of the bands.  
 
To approximately quantify the even-odd mixing, we define the superconducting gaps as follows:
\[
\tilde{\Delta}^{1,2}_{\uparrow, \downarrow}(\vk) = \Delta^{1,2}_{\uparrow, \downarrow}(\vk+\phi_{\vk}/2),
\]
on the spin-up Fermi surface, and
\[
\tilde{\Delta}^{2,1}_{\downarrow, \uparrow}(\vk) = \Delta^{2,1}_{\uparrow, \downarrow}(\vk-\phi_{\vk}/2),
\]
on the spin-down Fermi surface. At each Fermi wave vector \(\vk\) corresponding to the Fermi surface without SOC, we define \(\phi_{\vk}\) as representing half of the corresponding band splitting when SOC is introduced. Then, \(\vk \pm \phi_{\vk}\) approximately correspond to the Fermi wave vectors for the two spin-split bands, though the splitting is not necessarily symmetric.  

We can then approximately define the even- and odd-parity gaps as
\[
\Delta_{o/e}(\vk) = W_{1}(\vk)\tilde{\Delta}^{1,2}_{\uparrow \downarrow}(\vk) \pm W_{2}(\vk)\tilde{\Delta}^{2,1}_{\downarrow \uparrow}(\vk),
\]
where the weighting factor is given by
\[
W_{\tilde{m}}(\vk) = \frac{dl^{\tilde{m}}_{\vk}}{|\nabla\epsilon_{\tilde{m}}(\vk)|} \bigg/ \sum_{\vk} \frac{dl^{\tilde{m}}_{\vk}}{|\nabla\epsilon_{\tilde{m}}(\vk)|}.
\]
  To quantify the singlet-triplet mixing ratio, we define the parameter:
\[
\eta = \frac{1}{N_{\vk}}\sum_{\vk} \frac{|\Delta_e(\vk)| - |\Delta_o(\vk)|}{|\Delta_e(\vk)| + |\Delta_o(\vk)|},
\]
where \(N_{\vk}\) is a normalization factor. Thus within this formalism a pure even parity (odd parity) gap function corresponds to $\eta=1 (-1)$ respectively.

For an in-plane magnetic field on a 2D material, the presence of a spin-flip term implies that $S_z$ is no longer a good quantum number. As explained in the main text, for a field applied along the $x$ direction, the Zeeman term contribution in orbital space can be expressed as
\begin{align}
H_{B_x}=g\mu_B B_x\sum_{\vk,l} (c^{\dag}_{\vk l \uparrow}c_{\vk l \downarrow}+c^{\dag}_{\vk l\downarrow}c_{\vk l\uparrow}).
\end{align}

For only opposite-spin pairing in the presence of Ising spin–orbit coupling, the coupled set of linearized gap equations, Eq. (\ref{fullgapeqn}) can be written in matrix form as~\cite{roy2024unconventional}
\begin{widetext}
\begin{eqnarray}
    \begin{bmatrix}
 \Delta^{1,2}_{\uparrow,\downarrow}(\vk_{F_{1}}) \\
 \Delta^{2,1}_{\downarrow,\uparrow}(\vk_{F_{2}})
\end{bmatrix}
&=&-\sum_{\substack{\vk' }}
\begin{bmatrix}
 \left(\frac{I_{1}(\vk_{F_{1}}')V^{2,1,\downarrow,\uparrow}_{1,2,\uparrow,\downarrow}(\vk_{F_{1}},\vk_{F_{1}}')}{|\nabla\epsilon_{1,\uparrow}|_{\vk'=\vk_{F_{1}}'}}+
 \frac{I_{2}(\vk_{F_{2}}')V^{2,1,\downarrow,\uparrow}_{2,1,\downarrow,\uparrow}(\vk_{F_{1}},\vk_{F_{2}}')}{|\nabla\epsilon_{2,\downarrow}|_{\vk'=\vk_{F_{2}}'}}\right)
 & \left( \frac{I_{1}(\vk_{F_{2}}')V^{2,1,\downarrow,\uparrow}_{2,1,\downarrow,\uparrow}(\vk_{F_{1}},\vk_{F_{2}}')}{|\nabla\epsilon_{2,\downarrow}|_{\vk'=\vk_{F_{2}}'}}+\frac{I_{2}(\vk_{F_{1}}')V^{2,1,\downarrow,\uparrow}_{1,2,\uparrow,\downarrow}(\vk_{F_{1}},\vk_{F_{1}}')}{|\nabla\epsilon_{1,\uparrow}|_{\vk'=\vk_{F_{1}}'}}\right)
 \\
\left( \frac{I_{1}(\vk_{F_{1}}')V^{1,2,\uparrow,\downarrow}_{1,2,\uparrow,\downarrow}(\vk_{F_{2}},\vk_{F_{1}}')}{|\nabla\epsilon_{1,\uparrow}|_{\vk'=\vk_{F_{1}}'}}+
  \frac{I_{2}(\vk_{F_{2}}')V^{1,2,\uparrow,\downarrow}_{2,1,\downarrow,\uparrow}(\vk_{F_{2}},\vk_{F_{2}}')}{|\nabla\epsilon_{2,\downarrow}|_{\vk'=\vk_{F_{2}}'}}\right)
  & 
  \left(\frac{I_{1}(k_{F_{2}}')V^{1,2,\uparrow,\downarrow}_{2,1,\downarrow,\uparrow}(\vk_{F_{2}},\vk_{F_{2}}')}{|\nabla\epsilon_{2,\downarrow}|_{\vk'=\vk_{F_{2}}'}} +\frac{I_{2}(\vk_{F_{1}}')V^{1,2,\uparrow,\downarrow}_{1,2,\uparrow,\downarrow}(\vk_{F_{2}},\vk_{F_{1}}')}{|\nabla\epsilon_{1,\uparrow}|_{\vk'=\vk_{F_{1}}'}} \right)
  \label{eq:e_1}
\end{bmatrix}
\nonumber\\
&&\begin{bmatrix}
 \Delta^{1,2}_{\uparrow,\downarrow}(\vk_{F_{1}}') \\
 \Delta^{2,1}_{\downarrow,\uparrow}(\vk_{F_{2}}')
\end{bmatrix}
\end{eqnarray}
with the functions  $I_{1}(\vk)$ and $I_{2}(\vk)$ given in Eq. (\ref{eq_I12}).

\end{widetext}

\section{Calculation of DOS}

To calculate the density of states (DOS), we extrapolate the gap functions over the Brillouin zone. This is carried out using a radial basis function (RBF) interpolation, applied to the gap function defined on the Fermi surface and obtained from solving the linearized gap equation. The interpolation is further supplemented with a Gaussian envelope
\begin{align}
    \Delta(\vk)_{\text{BZ}} 
    = \Delta(\vk)_{\text{RBF}} 
    \times \exp\!\left(-\frac{\big(\epsilon_{m}(\vk)\big)^{2}}{\alpha^{2}}\right),
\end{align}
where $\epsilon_{m}(\vk)$ denotes the eigenenergy of band $m$. In the calculations, we set $\alpha = 0.1~\text{eV}$. For all parametrizations, the coefficients are chosen such that the maximum gap size is approximately
\[
\Delta_{\text{max}} \approx 0.005~\text{eV}.
\]

We then set up the Bogoliubov–de Gennes (BdG) Hamiltonian in matrix form:
\begin{align}
    \begin{pmatrix}
        \epsilon_{m}(\vk) & \Delta^{m}(\vk) \\
        \Delta^{m}(\vk)^* & -\epsilon_{m}(\vk)
    \end{pmatrix}
    \begin{pmatrix}
        u(\vk) \\[4pt]
        v(\vk)
    \end{pmatrix}
    =
    E_m(\vk)
    \begin{pmatrix}
        u(\vk) \\[4pt]
        v(\vk)
    \end{pmatrix}.
\end{align}
Here, $u(\vk)$ and $v(\vk)$ are the BdG coherence factors.  
The DOS on band $m$ at energy $\omega$ is obtained from the imaginary part of the following expression:
\begin{align}
\mathrm{N}_{m}(\omega) 
  &= - \, \mathrm{Im} \bigg[
     \frac{1}{\mathrm{N}_{\text{BZ}}} \, \frac{1}{\pi}
     \sum_{\vk} 
     \left(
        \frac{|u_{m}(\vk)|^{2}}
             {\omega - E_{m}(\vk) + i \eta} \right. \nonumber \\
  &\hspace{5.5em} \left.
      + \frac{|v_{m}(\vk)|^{2}}
             {\omega + E_{m}(\vk) + i \eta}
     \right)
   \bigg] .
\label{eq:dos}
\end{align}
where $\mathrm{N}_{\text{BZ}}$ is the number of $\vk$-points in the Brillouin zone used in the summation, and $\eta$ is a small positive broadening factor.

\bibliography{ref.bib}

\end{document}